\documentclass[12pt,english]{paper}
\usepackage[T1]{fontenc}
\usepackage[latin9]{inputenc}
\usepackage{babel}
 
\usepackage{graphicx}
\usepackage{caption}
\usepackage{subcaption}
\usepackage{xcolor}

\usepackage{amsmath}
\usepackage{amsfonts}
\usepackage{amsthm}
\usepackage{mathtools}
\usepackage{amssymb}
\usepackage{bbm}
 
\usepackage{color} 
\usepackage{hyperref}

\hypersetup{
  pdfborderstyle={/S/U/W 0}  
}

\usepackage{geometry}
\usepackage[normalem]{ulem}
\usepackage{natbib} 
\makeatletter

\theoremstyle{plain}

\theoremstyle{plain}

\theoremstyle{plain}
\newtheorem{proposition}{\protect\propositionname}
\theoremstyle{definition}
\newtheorem{defn}{\protect\definitionname}
\theoremstyle{plain}

\theoremstyle{plain}

\theoremstyle{plain}

\usepackage[colorinlistoftodos, obeyDraft]{todonotes} 

\providecommand{\definitionname}{Definition}
\providecommand{\lemmaname}{Lemma}
\providecommand{\propositionname}{Proposition}
\providecommand{\theoremname}{Theorem}
\providecommand{\assumptionname}{Assumption}
\providecommand{\corollaryname}{Corollary}
\providecommand{\remarkname}{Remark}

\makeatother

\begin{document}

\title{Arbitrageurs' profits, LVR, and sandwich attacks: batch trading as an AMM design response.\thanks{We are grateful to Felix Leupold and Martin K{\"o}ppelmann for initial discussions on batch trading on AMM that led to the writing of this paper, and to Eric Budish for his generous and detailed feedback. We also thank Haris Angelidakis, Andrea Barbon,  Agostino Capponi, Michele Fabi, Felix Henneke, Fernando Martinelli, Jason Milionis, Ciamac Moallemi, Andreas Park, Julien Prat, Tim Roughgarden, Anthony Lee Zhang, and participants at the Joint CEPR-Bocconi 2023 Conference ``The Future of Payments and Digital Assets'', Polytechnique de Paris-Oxford joint workshop on Blockchain and Decentralized Finance, Advances in Financial Technologies 2023 (Princeton University), ETHConomics 2023 at Devconnect Istanbul, Paris Dauphine Digital Days, Columbia CryptoEconomics (CCE) Workshop 2023 for numerous comments and suggestions.}}

\author{Andrea Canidio\thanks{Corresponding author; CoW Protocol; andrea@cow.fi} and Robin Fritsch\thanks{Matter labs}}

\maketitle

\noindent \today

\begin{abstract}
We study a novel automated market maker design:  the function maximizing AMM (FM-AMM). Our central assumption is that trades are batched before execution. Because of competition between arbitrageurs, the FM-AMM eliminates arbitrage profits (or LVR) and sandwich attacks, currently the two main problems in decentralized finance and blockchain design more broadly. We then consider 11 token pairs and use Binance price data to simulate the lower bound to the return of providing liquidity to an FM-AMM. Such a lower bound is, for the most part, slightly higher than the empirical returns of providing liquidity on Uniswap v3 (currently the dominant AMM).

\end{abstract}

\maketitle

\section{Introduction}

\todo[inline]{add application specific sequencer to the ways in which batching can be enforced}

Constant Function Automated Market Makers (CFAMMs) are the centerpiece of decentralized finance, handling daily trading volumes between USD 3 and 6 billion.\footnote{Unless otherwise noted, all relevant up-to-date statistics regarding decentralized finance can be found here: \url{https://defillama.com/dexs}.} Their popularity is largely due to their simplicity:  each CFAMM processes transactions sequentially as they appear on the blockchain, and the terms at which it is willing to trade depend exclusively on its reserves. Because each trade changes the composition of these reseves, different trades on the same CFAMM pay different prices even if they are added to the blockchain simultaneously (i.e., within the same block). The resulting pricing rule is \textit{discriminatory}, in the sense that trading a large amount in a single trade is equivalent to splitting the trade into smaller trades, each paying a different price.

This market mechanism has two well-recognized flaws. First, CFAMMs trade at a loss whenever there is a rebalancing event, in a way that is similar to the ``sniping'' of stale quotes in traditional financial markets (\citealp{foucault1999order}). When the underlying value of the assets changes, the first arbitrageur who trades with the CFAMM will bring its marginal price (i.e., the price of an arbitrarily small trade) in line with the equilibrium price. Because the price paid by the arbitrageur for this trade differs from the equilibrium price, this arbitrageur earns infra-marginal profits at the expense of the CFAMM's liquidity providers (LPs).  Second, traders are routinely exploited by attackers, most commonly via sandwich attacks in which an attacker front-runs a victim's swap with a swap in the same direction and then back-runs it with another swap in the opposite direction. Doing so allows the attacker to ``buy cheap'' and ``sell expensive'' while forcing the victim to trade at less favorable terms. 

These flaws, in turn, generate a race between arbitrageurs and between attackers in a way that is, again, akin to the high-frequency trading race in traditional finance. These races are particularly worrisome in blockchain-based finance because blockchains lack an objective notion of time: entities called \textit{validators} have full discretion on the ordering of pending transactions to form the next block, hence determining their order of execution. As a consequence,  arbitrageurs seeking to rebalance an AMM compete by paying validators to have their transactions included earlier in a block (similarly for attackers competing to sandwich a transaction). These payments are usually intermediated by entities called \textit{builders}, who are responsible for ordering transactions to maximize total profits (which are called Maximal Extractable Value, or MEV).  Because building blocks to extract maximum value is a specialized activity with increasing returns to scale, the builders' market is very concentrated: currently, the top three builders produce approximately  80\% of blocks added to the Ethereum blockchain. These builders exert great control over which transactions are included in the blockchain, therefore threatening the basic premise of blockchain as a permissionless and decentralized system. As it turns out, arbitrage profits and sandwich attacks constitute more than 95\%  of MEV transactions: without them, the benefit of delegating block construction to builders would be minimal.\footnote{Between October 2022 (when Ethereum changed its consensus protocol to proof of stake) to the end of 2023, validators earned approximately 410,000 ETH in MEV (approximately 1.2 billion USD at the current market price) For up-to-date statistics on total MEV, see \url{https://transparency.flashbots.net/}. 
On the challenges MEV poses for blockchain design, see the Ethereum foundation documentation (available here \url{https://ethereum.org/en/developers/docs/mev/}). For up-to-date statistics on the builders' market, see here \url{https://www.relayscan.io/overview?t=7d}. 
Several sources document how arbitrage profits and sandwich attacks constitute most of MEV; see, for example, Appendix 13 in \cite{heimbach2023ethereum}. Note that sandwich attacks contribute to the concentration in the builder market for an additional reason: as a defensive measure, a sophisticated user may pay a builder to include a transaction in a block privately, preventing sandwich attacks.  } 

This paper proposes an AMM design that eliminates arbitrage profits and sandwich attacks, and hence the vast majority of MEV. We start by building on existing literature (see the next section) in proposing an AMM design that processes transactions in \textit{batches} and uses \textit{uniform prices} (unlike traditional CFAMMs which process transactions sequentially and use discriminatory pricing). More precisely, we assume that all trades that reach the AMM during a period are batched together and executed at a price equal to the new marginal price on the AMM -- that is, the price of executing an additional small trade after the batch trades. We derive the trading function of such an AMM and show two equivalences. First, the condition specifying that the AMM trades at a price equal to the new marginal price can be interpreted as the first-order condition of a maximization problem. Hence, for given prices,  our  AMM trades to maximize a predetermined function of its reserves. For this reason, we call our design a \textit{function-maximizing} AMM, or \textit{FM-AMM}. Second, if the function is the product of the two liquidity reserves and FM-AMM does not charge any fee, its maximization implies that, for given prices, the value of the AMM's reserves is equally shared between the reserves assets. In other words, FM-AMM LPs run an equal-weight passive investment strategy (or, in the language of \citealp{milionis2022automated}, a \textit{rebalancing strategy}). Finally, we show that an FM-AMM does not satisfy path independence: traders can obtain a better price by splitting their trades into smaller orders, which is why batching is required.

Our main contribution is to study the behavior of an FM-AMM in the presence of arbitrageurs who have private information about the equilibrium prices (determined, for example, on some very liquid off-chain location). We show that competition between arbitrageurs guarantees that the batch always trades at the equilibrium price, and arbitrage profits are eliminated. Intuitively, if this were not the case, some arbitrageurs would want to trade with the batch and, by doing so, would push the price on the batch in line with the equilibrium.  FM-AMM, therefore, changes who benefits from arbitrageurs' competition: as already discussed, competition between CFAMM's arbitrageurs generates MEV and benefits validators; competition between FM-AMM's arbitrageurs instead benefits its liquidity providers who ``earn'' what would otherwise be MEV. An additional implication is that sandwich attacks are also eliminated: arbitrageurs will always act to remove deviations from the equilibrium price, making it impossible to manipulate the FM-AMM price. The intuition for our main result is therefore similar to that in \cite{budish2015high}, who show that batching trades eliminates latency races in traditional finance by changing the nature of competition between arbitrageurs. Our paper shows that the same solution can be adapted to eliminate the malicious reordering of transactions in decentralized finance.


We then propose a novel methodology to estimate LVR and the hypothetical benefit of an FM-AMM for liquidity providers. We empirically analyze the return of providing liquidity on Uniswap v3---currently the most important AMM with a market share of about 50\%---and compare it with a simulated counterfactual in which the same liquidity was provided to an FM-AMM receiving zero noise trading.\footnote{The absence of noise trading allows us to establish a lower bound for the return of providing liquidity to a more realistic FM-AMM with noise trading. We discuss in Section \ref{sec: empirical noise trading} how our empirical results change under different assumptions about FM-AMM noise trading volume. Note also that our theoretical model does consider the presence of noise traders.} 
We collect Binance price data for 11 token pairs from April to October 2023. The 11 token pairs we consider correspond to the highest-volume Uniswap v3 pools during the study period, excluding stable-coin to stable-coin pools and tokens not traded on Binance. For each token pair, our simulated FM-AMM charges the same fee as the corresponding Uniswap v3 pool. We assume that arbitrageurs can rebalance the FM-AMM pool to the corresponding Binance price. Because of the fees, arbitrageurs will not trade on the FM-AMM if the difference between the local and Binance prices is sufficiently small. Also, when arbitrageurs trade, they will trade less than required to rebalance the portfolio perfectly. Hence, a positive-fee FM-AMM always trades at the correct price (and arbitrageurs' profits are eliminated) but does not implement a rebalancing strategy as in \citealp{milionis2022automated}.

We compare the return on providing liquidity to this FM-AMM to the historical returns of providing liquidity to the corresponding Uniswap v3 pool.\footnote{Note that Uniswap v3 is characterized by  \textit{concentrated liquidity}: each LP provides liquidity over a price range and earns returns only if the price is within that range. In our comparison, we use the empirical distribution of liquidity and consider the return of an arbitrarily small non-concentrated liquidity position (i.e., a position over the entire price range $[0,\infty]$). For the FM-AMM, we also assume a liquidity position that is non-concentrated. Also, when calculating the return of providing liquidity to FM-AMM and Unsiwap v3, we assume that the fees earned are contributed back to the AMM as additional liquidity (which is not the default behavior of Unsiwap v3).  
} 
Our results generally support the benefit of providing liquidity to an FM-AMM: for 9 out of 11 token pairs, our simulated FM-AMM generates equal or higher returns than providing the same liquidity to Uniswap v3. The main exception is the MATIC-ETH pool, where Uniswap v3 outperforms our simulated FM-AMM by 0.65\%, which is also the largest absolute difference in returns across all token pairs. 
Section \ref{sec: empirical noise trading} extends the empirical analysis to allow noise trading on the simulated FM-AMM. The caveat is that, in our data, we do not know whether a given Uniswap transaction is noise trading, a sandwich attack, or a rebalancing by an arbitrageur. We, therefore, rely on previous work that measured the share of total volume attributable to each source across all Uniswap v3 pools. Assuming these shares are the same for our study pool, we can compute FM-AMM returns under different assumptions regarding its noise-trading volume relative to Uniswap. In particular, we show that also the FM-AMM MATIC-ETH pool outperforms the corresponding Uniswap v3 pool if its noise trading volume is at least half of Uniswap v3 MATIC-ETH pool's noise trading volume.


Precisely estimating the additional efficiency of changing the market mechanism from a traditional AMM to an FM-AMM is beyond the scope of the paper. We can, however,
provide some back-of-the-envelope estimations. First, there is the direct benefit of eliminating sandwich attacks. On Ethereum alone, the cost generated by these attacks was, on average, USD 500.000 daily in December 2023,\footnote{For up-to-date statistics, see \url{https://eigenphi.io/mev/ethereum/sandwich}.} or between 2.5 and 5 basis points of daily DEX volume on Ethereum. For reference, \cite{aquilina2022quantifying} finds that latency arbitrage in traditional financial markets imposes a roughly 0.5 basis point tax on trading. Second, eliminating arbitrage profits increases the return of providing liquidity to an AMM. For a ballpark estimate of the size of this increase, we use the results in  \cite{milionis2022automated} to estimate arbitrage profits from price volatility. We calculate that, in 2023, eliminating arbitrage profits would have added between 5 and 7 percentage points to the return of liquidity providers in ETH/stable and BTC/stable pools with unconcentrated liquidity, and even more for liquidity providers of other tokens (usually more volatile than ETH or BTC) and for pools with concentrated liquidity. Because there are approximately USD 10 billion in non-stable-to-stable AMM pools, eliminating arbitrage profits would have added between USD 0.5 and 1 billion to LPs profits in 2023. 


To conclude this section, we want to stress that FM-AMM may, in the future, be relevant also to traditional finance. There is a growing appreciation of the potential benefit of exchanging \textit{traditional} financial assets using blockchain-based AMMs, primarily due to increased efficiency (i.e., the simultaneity of clearing and settlement) and security. A case in point is the BIS Mariana project, testing the use of an AMM for exchanging Central Bank Digital Currencies\footnote{See \url{https://www.bis.org/publ/othp75.htm}.} or the recent decision by Franklin Templeton (one of the world's largest fund managers) to launch a blockchain-based money-market fund.\footnote{See \url{https://www.franklintempleton.co.uk/press-releases/news-room/2023/franklin-templeton-money-market-fund-launches-on-polygon-blockchain}). }  In addition to these benefits, \cite{malinova2023learning} calculate that exchanging equities using a traditional AMM  could save U.S. investors about 30\% of annual transaction costs, due to improved risk sharing and repurposing of idle capital. By eliminating sandwich attacks and arbitrage profits, the FM-AMM may help realize these benefits. A final observation is that at least two companies started developing an FM-AMM by taking direct inspiration from earlier versions of our paper, showing that some industry practitioners believe that FM-AMM may have the benefits that our theory and empirical analysis predict.\footnote{The first company is CoW Protocol, the employer of one of the authors of this paper. CoW Protocol collects trades into a batch and then runs an auction between entities called ``solvers'' who compete to provide the best possible execution of these trades by accessing public AMMs,  private liquidity sources, or finding coincidence of wants (CoW) directly on the batch. It currently intermediates more than 2 billion USD monthly. CoW Protocol recently launched an AMM inspired by FM-AMM (see here \url{https://cow.fi/cow-amm}). The second is a  startup called ``sorella labs'' (see here \url{https://twitter.com/SorellaLabs}).}


\paragraph{Relevant literature}

Several authors argued that AMM's design allows arbitrageurs to profit at the expense of LPs. \cite{aoyagi2020liquidity}, \cite{capponi2021adoption},  \cite{milionis2022automated}, and \cite{milionis2023automated} provide theoretical models that illustrate this possibility. In particular,  \cite{milionis2022automated} consider a continuous time model with zero fees and derive a closed-form formula to measure LPs returns and the cost they face when trading with informed arbitrageurs (which they call loss-vs-rebalancing or LVR). \cite{milionis2023automated} extend the analysis to blocks added at discrete time intervals and strictly positive trading fees. They use the term \textit{arbitrageur profits} to indicate LPs losses, a term we adopt because we also assume that blocks are added at discrete time intervals and there may be non-zero fees.  \cite{capponi2021adoption} show that arbitrageurs can exploit LPs even without asymmetric information. The reason is that the first arbitrageur trading with the pool can profit at the expense of \textit{several} LPs. Hence, this arbitrageur's profits are larger than any individual LPs loss and can always outspend (and trade before) any LP who may want to withdraw their liquidity. 
Also related are \cite{lehar2021decentralized} and \cite{foley2022can}, who argue that liquidity provision to AMMs is strategic: the size of liquidity pools is smaller when arbitrageurs' profits are higher. The intuition is that when there is a rebalancing event, the loss to arbitrageurs \textit{per unit of liquidity} is independent of the size of the liquidity reserves. At the same time, the size of the liquidity reserves determines the fraction of the revenues from noise traders earned by each unit of liquidity. Hence, the endogenous response of liquidity providers may explain why we find that fees from noise traders are approximately equal to arbitrageurs' profits in Uniswap v3. 


A second important limitation of CFAMMs is that they enable sandwich attacks (see \citealp{park2022conceptual}). These attacks are quantitatively relevant. For example, \cite{torres2021frontrunner} collected on-chain data from the inception of Ethereum (July 30, 2015) until November 21, 2020, and estimated that sandwich attacks generated USD 13.9M in profits.  \cite{qin2022quantifying}  consider a later period (from December 1, 2018 to August 5, 2021) and find that sandwich attacks generated  USD 174.34M in profits. As already mentioned, the data provider \url{https://eigenphi.io/} on Ethereum measures in real-time the cost caused by sandwich attacks (measured as profits to the attacker plus the payment to the validator for inclusion in the block)  and currently reports approximately USD 17 M for December 2023. Because an FM-AMM eliminates these attacks, our paper is part of the small but growing literature proposing mechanisms to prevent malicious re-ordering of transactions by changing the design of blockchain applications (vs. changing blockchain infrastructure); see \cite{breidenbach2018enter},  \cite{gans2022solomonic}, \citeauthor{canidio2023commitment} (forthcoming),  \cite{ferreira2023credible}.

Several initial discussions on designing ``surplus maximizing'' or ``surplus capturing'' AMMs occurred informally on blog and forum posts (see \citealp{leupold2022cow}, \citealp{josojo2022mev}, \citealp{dellapenna2022mev}). The idea of an AMM that accepts trades from a batch is already in
\cite{goyal2022batch} and \cite{johnson2023concave}. In particular,  \cite{goyal2022batch} derive several possible trading rules for such AMM, one of which corresponds to the FM-AMM (trading rule U). The FM-AMM pricing function is also studied in  \cite{park2022conceptual}, who calls it \textit{uniform pricing} AMM. He also shows that such an AMM violates path independence because it generates the incentives to split trades.\footnote{In an earlier paper, \cite{forgy2021family} studies a family of AMM pricing functions, one of which corresponds to the FM-AMM. However, they do not discuss the issue of trade splitting or batching.} Relative to these works, our main contribution is to study FM-AMM in a context with privately informed arbitrageurs. Competition between these arbitrageurs guarantees that the FM-AMM always trades at the ``correct'' equilibrium price, eliminating arbitrage profits and sandwitch attacks.  
Finally, \cite{schlegel2022axioms} study AMM from an axiomatic viewpoint. In particular, they discuss path independence, which FM-AMMs violate. 

\todo[inline]{with respect to the pro-rata games paper: "Pro-rata" means that there is an equilibrium price vector - like in 80\% of all econ papers dealing with the existence of prices. So I don't think we should relate to their game more than we should relate to 80\% of the econ literature. Furthermore, their analysis of the case with arbitrageurs is incorrect: they say that there are $n$ arbitrageurs, but then in the mathematical expressions they present, they only have 2. By the way, if they had the right formula, by taking $n$ to infinity, they would have the same result: arbitrageurs trade on the AMM at the outside price  }

Several authors studied whether providing liquidity on AMMs, and in particular, on Uniswap, is profitable; see \cite{heimbach2021behavior}, \cite{loesch2021impermanent}, \cite{heimbach2022risks}. The main difference between these papers and ours is that we compare Uniswap LP returns to a different benchmark (here, the return of contributing liquidity to a positive-fee FM-AMM, in those papers, a holding strategy). In this respect, our strategy is similar to the empirical analysis in \cite{milionis2022automated}, who find that for the ETH-USDC Uniswap v2 pool, the losses to arbitrageurs are smaller than the revenues from noise traders. The main difference with our analysis is in the benchmark against which we compare Uniswap LPs returns. In \cite{milionis2022automated}, the benchmark is a rebalancing strategy, equivalent to providing liquidity to a zero-fee FM-AMM. Instead, we consider an FM-AMM with the same fee as the corresponding Uniswap pool. This implies that if the reference Uniswap pool is not rebalanced because its fees are too high relative to the price movement, the same applies to our benchmark FM-AMM. Another difference is in the implementation of the benchmark, which, in our case, is not a theoretical construct but an alternative AMM design. Finally, we study several Uniswap v3 pools.

Our paper is closely related to the literature studying batch auctions in traditional finance. In particular, the intuition for our main result is closely related to \cite{budish2015high}, who study the batching of trades in the context of traditional finance as a way to mitigate the high-frequency-trading (HFT) arms race and protect regular (or slow) traders. The main result is that when trading happens in continuous time, arbitrageurs compete on speed. When traders are instead batched (that is, collected over a given time interval and then settled), arbitrageurs compete on price because the priority of execution within the batch is given based on price. The intuition in our model is similar, although competition between arbitrageurs on the batch is rather in quantity than in price: if the price on an FM-AMM differs from the equilibrium price, competing arbitrageurs will submit additional trades to exploit the available arbitrage opportunity, but by doing so, they push the price on the FM-AMM in line with the equilibrium. See also \cite{aquilina2022quantifying}, who estimates the cost of high-frequency trading races. Also relevant is an older literature comparing batch trading with continuous time double auction and showing that batch trading reduces the inefficiencies due to the presence of informed traders (\citealp{madhavan1992trading}) and reduces the cost of intermediation by allowing traders to trade with each other directly (\citealp{nicholas1995electronic}).   Finally, our work is in the spirit of literature studying how to design real-world markets (as in \citealp{roth2002economist}). In particular, \cite{roth1997turnaround} studies the benefit of batch processing over sequential processing but focuses on a very different market.




We conclude by noting that an FM-AMM is also an oracle: it exploits competition between arbitrageurs to reveal on-chain the price at which these arbitrageurs can trade off-chain. It is, therefore, related to the problem of Oracle design (as discussed, for example, by \citealp{chainlink}).

\paragraph{Outline} 
The remainder of the paper is organized as follows. In Section \ref{Sec: simple model}, we introduce the FM-AMM in its simplest form with a product function and zero fees. In Section \ref{Sec: extensions}, we discuss several extensions, including fees and the fact that an FM-AMM violates path dependence and hence requires batching. In Section \ref{sec: the model}, we consider the behavior of an FM-AMM in the equilibrium of a game with informed arbitrageurs and noise traders. Section \ref{sec: empirical analysis} contains the empirical analysis. Section \ref{sec: attack} discusses the threat model, that is, what happen when the entity that manages the batching process is malicious. The last section concludes. 

\section{The function-maximizing AMM}\label{Sec: simple model}

In this section, we first introduce the main concepts of interest using a simple constant-product function (both for the CFAMM and the FM-AMM), no fees, and keeping formalities to the minimum. In the next section, we generalize our definitions and results and introduce additional elements.

\begin{figure}[t]
    \centering
     \begin{tikzpicture}[scale=1.7, every node/.style={scale=.7}]
    \draw[->] (0,0) -- (4.5,0) node[below right] {ETH};
    \draw[->] (0,0) -- (0,4.5) node[above left] {DAI};
    \draw[domain=0.25:4,smooth,variable=\x,blue] plot ({\x},{1/\x}) ;
   
 \draw[dashed] (0,2+.5*4)-- (.5,2) --(2/4+.5,0);

 \draw[dotted](0,2)node[left]{$Y+x\cdot p(x)$} -- (.5,2) -- (.5,0)   node[below]{$X - x$};
 \draw[dotted] (0,.5)node[left]{$Y$} -- (2,.5) -- (2,0) node[below]{$X$};

\draw (2/4+.3,0) arc[start angle=0, end angle=-40, radius=-.5];
    \node at (2/4+.15,.2)  {$-\hat p$};

\draw (2.2,0) arc[start angle=0, end angle=-28, radius=-.5];
    \node at (1.7,.2)  {$- p^{CPAMM}(x)$};

\draw(0,2.5)--(2.5,0);

\end{tikzpicture}
    \caption{Initially, the liquidity reserves of the CPAMM are $Y$ and $X$. A trader then purchases $x$ ETH at an average price $p(x)$. Note that, after the trade, the marginal price on the CPAMM (that is, the price for an arbitrarily small trade) is $\hat p \neq p^{CPAMM}(x).$  \label{fig:1}}
   
\end{figure}
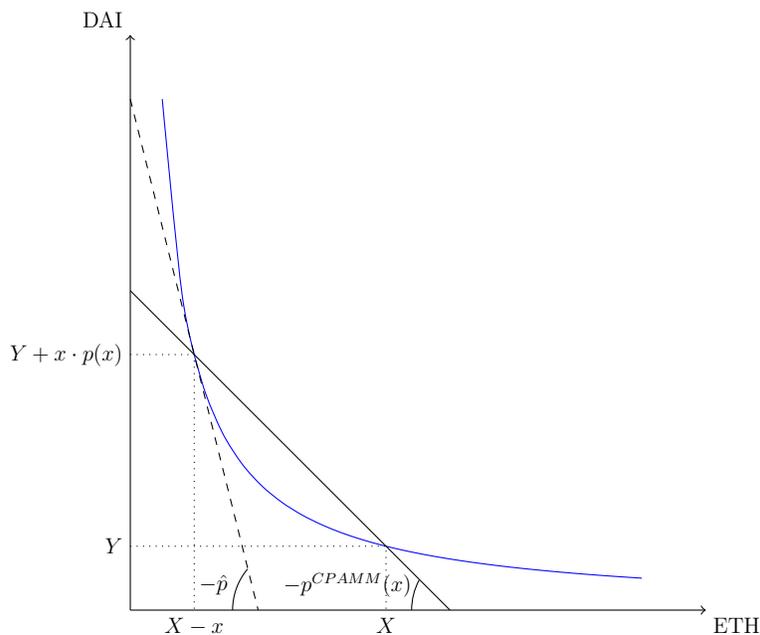

As a preliminary step, we derive the trading function of a constant product AMM, the simplest and most common type of CFAMM. Suppose that there are two assets, ETH and DAI. A constant product AMM (CPAMM) is willing to trade ETH for DAI (or vice versa) as long as the product of its liquidity reserves remains constant (see Figure  \ref{fig:1} for an illustration). Call $Y$ and $X$ its initial liquidity reserves in DAI and ETH, respectively, and $p^{CPAMM} (x)$ the average price at which the CPAMM is willing to trade $x$ ETH, where $x>0$ means that CPAMM is selling ETH while $x<0$ means that the CPAMM is buying ETH. For the product of the liquidity reserves to be constant, it must be that 
\[
Y\cdot X = (Y+p^{CPAMM} (x) \cdot x) (X -x)
\]
or
\[
p^{CPAMM} (x)=\frac{Y}{X-x}.
\]
Note that the marginal price of a CPAMM (i.e., the price to trade an arbitrarily small amount) is equal to the ratio of its liquidity reserves. The key observation is that, in a CPAMM, a trader willing to trade $x$ pays a price different from the marginal price after the trade. This is precisely the reason why arbitrageurs can exploit a CPAMM:  an arbitrageur who trades with the CPAMM to bring its marginal price in line with some exogenously determined equilibrium price does so at an advantageous price (and hence makes a profit at the expense of the CPAMM LPs).

Instead, in the introduction, we defined an FM-AMM as an AMM in which, for every trade, the average price equals the marginal price after the trade -- a property we call \emph{clearing-price consistency}. For ease of comparison with the CPAMM described earlier, suppose that the FM-AMM function is the product of the two liquidity reserves. If its marginal price is, again, the ratio of its liquidity reserves, then the AMM is clearing-price consistent if and only if its price function $p(x)$ is
\begin{equation*}
    p(x) =  \frac{Y +  x \cdot p(x)}{X- x},
\end{equation*}
where the RHS of the above expression is the ratio of the two liquidity reserves after the trade. Solving for $p(x)$ yields:
\begin{equation*}
    p^{FM-AMM} ( x) \equiv p( x) = \frac{Y}{X-2 x},
\end{equation*}
which implies that the FM-AMM's marginal price is, indeed, the ratio of the liquidity reserves. Hence, a given trade on the FM-AMM generates twice the price impact than the same trade on the traditional CPAMM (cf. the expression for $p^{CPAMM} (x)$).

\begin{figure}[ht]
    \centering
     \begin{tikzpicture}[scale=1.5, every node/.style={scale=.7}]

    \draw[->] (0,0) -- (4.5,0) node[below right] {ETH};
    \draw[->] (0,0) -- (0,6) node[above left] {DAI};
    \draw[domain=0.25:4,smooth,variable=\x,blue] plot ({\x},{1/\x}) ;
   
 \draw[dashed] (0,2+.5*4)-- (.5,2) --(2/4+.5,0);

 \draw[dotted] (0,.5)node[left]{$Y$} -- (2,.5) -- (2,0) node[below]{$X$};

\draw(2-.08,0) arc[start angle=0, end angle=-40, radius=-.5];
    \node at (2-.4,.2)  {$-p(x)$};


\draw(.8,{(2+.5*4)/(2/4+.5)*(2.13-.8)})--(2.13,0);

\draw[blue, ->](3,.4)--(3.3,1.3);

\draw[blue, ->](1.1,1.1)--(2,2);

\draw[blue, ->](.5,3)--(1.2,3.2);

  \draw[domain=.8:4,smooth,variable=\x,blue] plot ({\x},{4.55/\x}) ;

\end{tikzpicture}
    \caption{On an FM-AMM, the price at which a given trade $x$ is executed equals the marginal price after the trade is executed.  This implies that an FM-AMM ``moves up'' the curve with each trade.}
    \label{fig:2}
\end{figure}
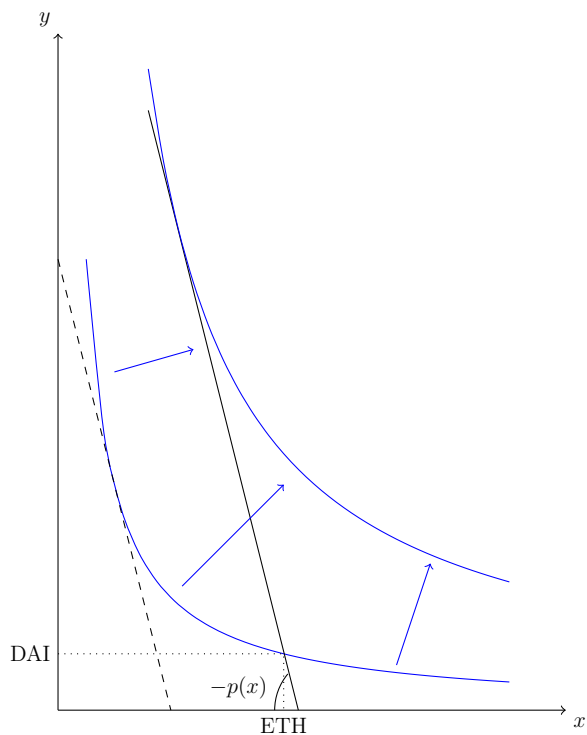

Interestingly, an FM-AMM can also be seen as a price-taking agent maximizing an objective function. If its objective function is the product of the two liquidity reserves, then for a given price $p$ the FM-AMM supplies $x$ ETH by solving the following problem:
\[
x^{FM-AMM}(p)=\mbox{argmax}_{ x} \left\lbrace  (X- x)(Y+p\cdot  x) \right\rbrace.
\]
It is easy to check that the FM-AMM supply function is:
\[
 x^{FM-AMM}(p) = \frac{1}{2} \left(X -\frac{Y}{p} \right).
\]
Hence, to purchase $  x$ ETH on the FM-AMM, the price  needs to be, again:
\[
p^{FM-AMM}( x) = \frac{Y}{X-2 x}.
\]
It follows that,  whereas a traditional CPAMM always trades along the same curve given by $Y \cdot X $, the FM-AMM trades as to be on the highest possible curve. With some approximation, we can see an FM-AMM as a traditional CPAMM in which additional liquidity is added with each trade. See Figure \ref{fig:2} for an illustration. 

A final observation is that the FM-AMM's trading function is equivalent to
\[
p\cdot (X- x^{FM-AMM}(p))=Y+p \cdot x^{FM-AMM}(p).
\]
In other words, for a given $p$, the values of the two liquidity reserves are equal after the trade. Therefore, an FM-AMM with product function trades to implement a passive investment strategy, in which the total value of the two reserves is equally split between the two assets (that is, a passive investment strategy with weights $1/2$, $1/2$). It is easy to check that the FM-AMM can implement any passive investment strategy with fixed weights $(\alpha, 1-\alpha)$ by specifying the objective function as $(X- x)^\alpha (Y+p \cdot x)^{1-\alpha}$.

\section{Additional considerations}\label{Sec: extensions}
\subsection{Generalization of definitions and results}

We now generalize our results. We define an AMM as an entity that accepts or rejects trades based on a pre-set rule. Such a rule can be derived from the AMMs liquidity reserves $(Y,X) \in \mathbb{R}^2_+$ and the AMM function $\Psi: \mathbb{R}^2_+ \to \mathbb{R}$. We assume that the AMM function is continuous, it is such that $\Psi(Y,0)=\Psi(0,X)=\Psi(0,0)$ for all $Y$ $X$, that it is strictly increasing in both its arguments whenever $Y>0$ and $X>0$, and that it is strictly quasiconcave. The difference between different types of AMMs is how the function $\Psi(.,.)$ and the liquidity reserves $(Y,X)$ determine what trades will be accepted and rejected by the AMM. 

\begin{defn}[Constant Function Automated Market Maker]
For given liquidity reserves $(Y,X)$ and function $\Psi: \mathbb{R}^2_+ \to \mathbb{R}$, a constant function automated market maker (CFAMM) is willing to trade $  x$ for $ y=p(x) \cdot x$ if and only if
\begin{equation*}
\Psi(Y + p(x) \cdot  x,X-x) = \Psi(Y,X).
\end{equation*}
\end{defn}

Our first goal is to define an AMM  that is \textit{clearing-price-consistent} in the sense that, for every trade, the average price of the trade equals the marginal price after the trade.

\begin{defn}[Clearing-Price Consistent AMM]
For given liquidity reserves $(Y,X)$ and function $\Psi: \mathbb{R}^2_+ \to \mathbb{R}$, let $$p_\Psi^{margin}(Y,X) =  \frac{\frac{\partial \Psi(Y,X) }{\partial X}}{\frac{\partial \Psi(Y,X) }{\partial Y}}$$ 
be the marginal price of the AMM for reserves $(Y,X)$. 
A clearing-price consistent AMM is willing to trade $  x$ for $  y=p(x) x$ if and only if
\begin{equation}\label{eq:def_cpc_amm}
    p( x) = p_\Psi^{margin}\left(Y+p( x) x, X- x \right).
\end{equation}

\end{defn}
Note that, given our assumptions on  $\Psi(.,.)$,  whenever $Y>0$ and $X>0$, the marginal price is strictly increasing in the first argument and strictly decreasing in the second argument,  converges to zero as $X \to \infty$ or $Y \to 0$, and to infinity as $X \to 0$ or $Y \to \infty$. 


We also define a \textit{function-maximizing AMM (FM-AMM)} that maximizes the objective function instead of keeping it constant:

\begin{defn}[Function-Maximizing AMM]
For given liquidity reserves $(Y,X)$ and function $\Psi: \mathbb{R}^2_+ \to \mathbb{R}$, a function-maximizing AMM is willing to trade $   x$ for $   y=p(  x) \cdot   x$ if and only if $p(  x)=  x^{-1}(p)$, where
\begin{equation}\label{eq:sm_amm}
      x(p) := \mbox{argmax}_{  x} \left\lbrace   \Psi\left( Y+p \cdot   x, X-  x \right)   \right\rbrace.
\end{equation}
\end{defn}

The next proposition establishes the equivalence between clearing-price-consistent and function-maximizing AMMs.

\begin{proposition}\label{prop:equivalence}
For given liquidity reserves $(Y,X)$ and function $\Psi: \mathbb{R}^2_+ \to \mathbb{R}$, an AMM is function maximizing if and only if it is clearing-price consistent. 
\end{proposition}
\begin{proof}
    Under our assumptions, solving \eqref{eq:sm_amm} is equivalent to satisfying the first-order condition, which is equivalent to \eqref{eq:def_cpc_amm}.
\end{proof}

\subsection{Path-dependence (or why batching trades is necessary)}\label{sec: path dependence}

CFAMMs are path-\textit{independent}: splitting a trade into multiple parts and executing them sequentially does not change the average price of the trade. This property does not hold for an FM-AMM because traders can get better prices by splitting their trade. In fact, by splitting their trade into arbitrarily small parts, traders pay approximately the same price as on the corresponding CFAMM. This is why an FM-AMM's trading function can be implemented only if trades are batched.

To see this, note trading $x$ on an FM-AMM with product function changes the reserves as follows:
\begin{equation*}
    \left(Y, X\right) \quad\rightarrow\quad \left(Y\left(\frac{X-  x}{X-2  x}\right), X-  x\right).
\end{equation*}
If instead we split the trade into smaller parts $\sum_{i=1}^n   x_i =   x$ and execute them sequentially, the FM-AMM reserves will change to
\begin{equation*}
    \left(Y \prod_{i=1}^n \frac{(X-\sum_{j=1}^{i-1}  x_j)-  x_i}{(X-\sum_{j=1}^{i-1}  x_j)-2  x_i}, X - \sum_{i=1}^n   x_i\right).
\end{equation*}
Setting $  x_i=\frac{1}{n}  x$ and letting $n\to\infty$ leads to the DAI reserves after the trade being
\begin{equation*}
    \lim_{n\to\infty} Y \frac{X-\frac{1}{n}  x}{X-\frac{n+1}{n}  x} =  Y \frac{X}{X-  x},
\end{equation*}
which equals the DAI reserve of a CPAMM after a trade $  x$. Hence, to have an FM-AMM, it is necessary to prevent splitting orders by imposing the batching of trades.\footnote{It is still possible that trades are split across batches. Whether this is profitable depends on who else is trading with the FM-AMM, which is a problem with the study in the next section.}
\todo[inline]{generalize this to an FM-AMM derived from an arbitrary trading function $\Psi$. If we ever add this, we could use a second-order taylor expansion. The first term is the same between CFAMM and FM-AMM, while the second differs. However, as trades are split into smaller and smaller chunks, the second term vanishes to zero (perhaps!).}


\subsection{Batching}\label{sec: batching}


In what follows, we assume that the FM-AMM enforces batching by collecting intentions to trade off-chain and settling them on-chain each block. Trades in opposite directions are settled peer-to-peer, and the excess is settled on the FM-AMM as a single trade. Importantly, all trades belonging to the same batch face the same prices before fees (the next section discusses the role of fees).\footnote{This process is modeled around CoW Protocol (\url{www.cow.fi}). CoW Protocol collects intentions to trade off-chain and executes them as a batch. Cow Protocol enforces uniform clearing prices so that all traders in the same batch face the same prices. Also, all intention-to-trade collected by CoW Swaps for inclusion in its batch are visible to all market participants.} We can, therefore, think of batching as an off-chain component of the AMM, together with, for example, its UI.

The fact that batching is done off-chain introduces a trust assumption, which we discuss in Section \ref{sec: attack}. However, it is important to note that, in the future, there may be other ways to enforce batching, some of which remove this off-chain component. For example, if the FM-AMM is built on Ethereum, batching could be enforced by leveraging proposer-builder separation (or PBS). In PBS, block builders (entities that assemble transactions in a block that are then forwarded to a proposer for inclusion in the blockchain) could compute the net trades that will reach the FM-AMM during that block and include a message at the beginning of the block announcing this value. The FM-AMM, then, uses this message to compute the price at which all trades will be executed. If the proposer's announcement turns out to be correct at the end of the block, the FM-AMM will reward the builder (punishments can also be introduced if the block builder report is incorrect, see \citealp{leupold2022cow}). An alternative is to use an application-specific layer 2 blockchain (or app chain): a blockchain designed to operate one specific application, and that settles on an underlying blockchain (for example, Ethereum) at regular intervals. How to do batching in an app-chain is an area of active research.


\subsection{Fees}\label{sec:fees}

To study the possibility that FM-AMM charges fees, we first consider the case where all orders on the batch have the same sign (that is, all orders are either buy or sell). We then extend the analysis to the case in which there could be both buy and sell orders on the batch, and hence the FM-AMM has both a buy and a sell price. For ease of comparison with Uniswap (the most important and liquid CFAMM), we assume that the fee is paid in the sell tokens, that is, the input token from the AMM perspective. The fee collected accrues within FM-AMM as additional liquidity.\footnote{FM-AMM, therefore, behaves similarly to Uniswap V1 and V2. On Unsiwap V3, however, the default behavior is that fees accrue outside the AMM. Nonetheless, in our empirical estimation of the return of providing liquidity to Uniswap V3, we will assume that the fees earned are returned to the AMM as additional liquidity.}

\paragraph{Either buy or sell orders on the batch.} 
Suppose there is a fee $\tau \in (0,1)$. If $x<0$  (i.e., the batch sells ETH to the FM-AMM), then a fraction of the sell amount is captured as fees, and the rest is traded on the FM-AMM. The traders on the batch therefore receive $(1-\tau) \cdot p^{FM-AMM}(  x(1-\tau))$ DAI for each ETH that they sell to the FM-AMM. If instead  $x>0$ (i.e., the batch purchases ETH from the FM-AMM), then again a fraction of the sell amount (in this case, DAI) is captured by the FM-AMM as fees. It follows that the traders on the batch pay $  p^{FM-AMM}(  x)/(1-\tau)$ DAI for each ETH that they buy from the FM-AMM.  For $x\neq 0$, we can therefore define the FM-AMM \textit{effective price} as
\begin{equation}\label{eq: effective prices}
\tilde p(  x, \tau) \equiv \begin{cases}  \frac{ p^{FM-AMM}(  x)}{1-\tau}=\frac{ Y}{(1-\tau)(X-2  x)} &\mbox{ if }   x> 0 \\  (1-\tau) \cdot p^{FM-AMM}(  x(1-\tau))=\frac{Y }{\frac{X}{(1-\tau)}-2  x} &\mbox{ if }   x < 0 \\ 
\end{cases}
\end{equation}
The terms $X/(1-\tau)$ and $Y/(1-\tau)$ have an intuitive interpretation: the fee causes the FM-AMM to behave as if it had more of the token that traders want to sell to the FM-AMM. This also implies that the fee $\tau$ affects the elasticity of the effective price to the size of the trade $|  x|$, with a higher fee implying a larger price impact.

\begin{figure}[ht]
    \centering
     \begin{tikzpicture}[scale=1.8, every node/.style={scale=.7}]
   \def\t{.5}; 
   \def\QUSD{1.15}; 
   \def\QE{1.15}; 
   
    \draw[->] (0,0) -- (6.5,0) node[below right] {ETH};
    \draw[->] (0,0) -- (0,6) node[above left] {DAI};
    \draw[domain=0.3:4,smooth,variable=\x,dotted] plot ({\x},{\QUSD*\QE/\x}) ;

 \draw[dotted] (0,\QUSD)node[left]{$Y$} -- (\QE,\QUSD) -- (\QE,0) node[below]{$X$};
 \draw[dotted] (4,\QUSD) -- (\QE,\QUSD) -- (\QE,4);

     \draw[domain=.45:1.15,smooth,variable=\x, blue] plot ({\x},{3.1/\x-(\t/(1-\t))*\QUSD}) ; 
      \draw[domain=.4:1.15,smooth,variable=\x, blue] plot ({\x},{2.65/\x-(\t/(1-\t))*\QUSD}) ; 
     
     \draw[domain=1.55:6.5,smooth,variable=\x, blue] plot ({\x},{(3.1/(\x+\QE*(\t/(1-\t)))}) ;

        \draw[domain=1.18:6.5,smooth,variable=\x, blue] plot ({\x},{(2.63/(\x+\QE*(\t/(1-\t)))}) ;
 
 \draw (0,1.4) --(6.25,0);

\draw(5,0) arc[start angle=0, end angle=-30, radius=-.5];
    \node at (4.6,0.15) {$-\tilde p(  x, \tau)$};

    \draw[dotted] (0,.8)node[left]{$Y +  x \tilde p(  x, \tau)$} -- (2.6,.8) -- (2.6,0) node[below]{$X-  x$};

\end{tikzpicture}
    \caption{A positive-fee FM-AMM moves up the curve: effective price when the batch trades $  x<0$, and there are no buy orders (in blue, the FM-AMM level curves for given $Y$ and $X$).}
    \label{fig:function-max-with-fees}
\end{figure}
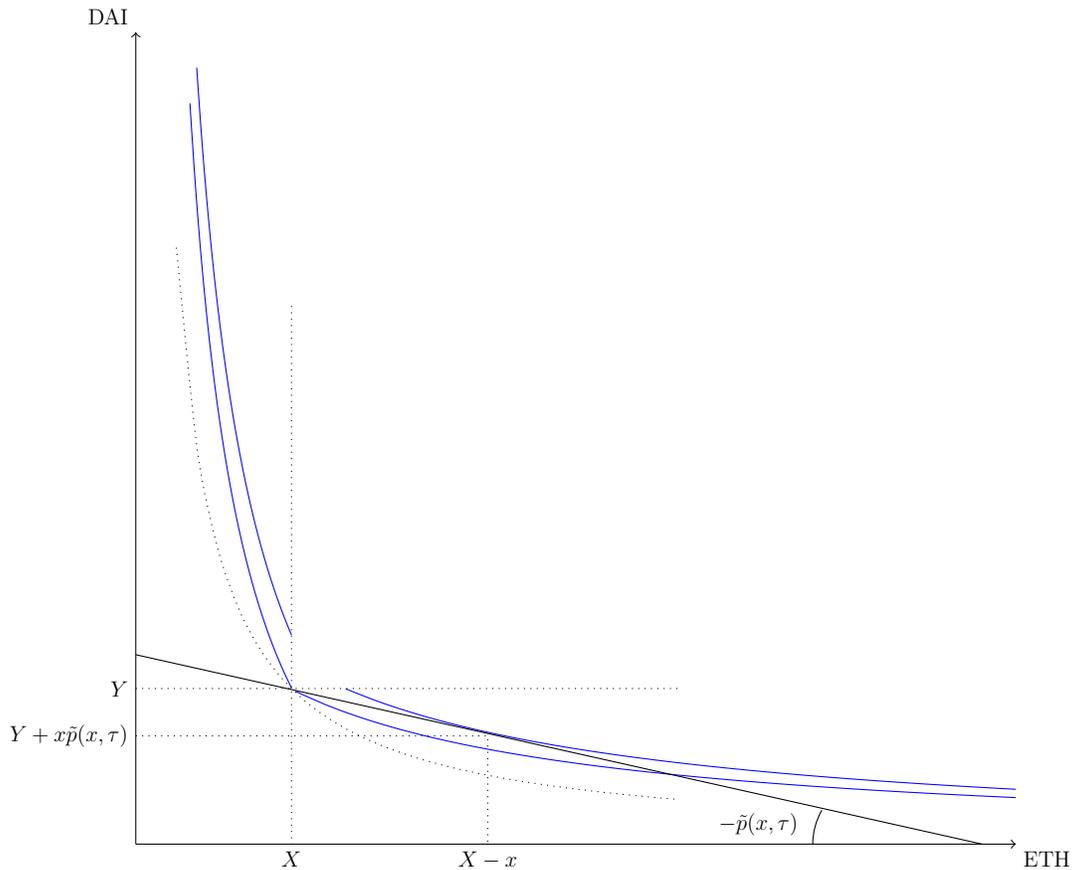

A final observation is that a positive-fee FM-AMM remains a function maximizing AMM, but the objective of the maximization depends on the sign of the trade. We can therefore write $\tilde p(  x, \tau) =   x^{-1}(p, \tau) $ and
\[
  x(p, \tau)=\mbox{argmax}_{  x} \left\lbrace U(  x,p, \tau) \right\rbrace
\]
where
\begin{equation}\label{eq: maximand with fees}
U(  x,p, \tau) =\begin{cases} 
 \left(X -  x\right) \cdot \left(\frac{Y}{1-\tau} +p\cdot   x \right)        &\mbox{ if }   x\geq 0 \\
 \left(\frac{X}{1-\tau} -  x\right) \cdot \left(Y +p \cdot   x \right)  &\mbox{ if }   x\leq 0 \\
 \end{cases}
\end{equation}
See Figure \ref{fig:function-max-with-fees}.

\paragraph{Both buy and sell orders on the batch.} 

Suppose now that there are both buy and sell orders on the batch. In this case, some orders are settled peer-to-peer, while other orders may be settled on the FM-AMM.  All orders will have the same price before fees but different buy and sell prices after fees.

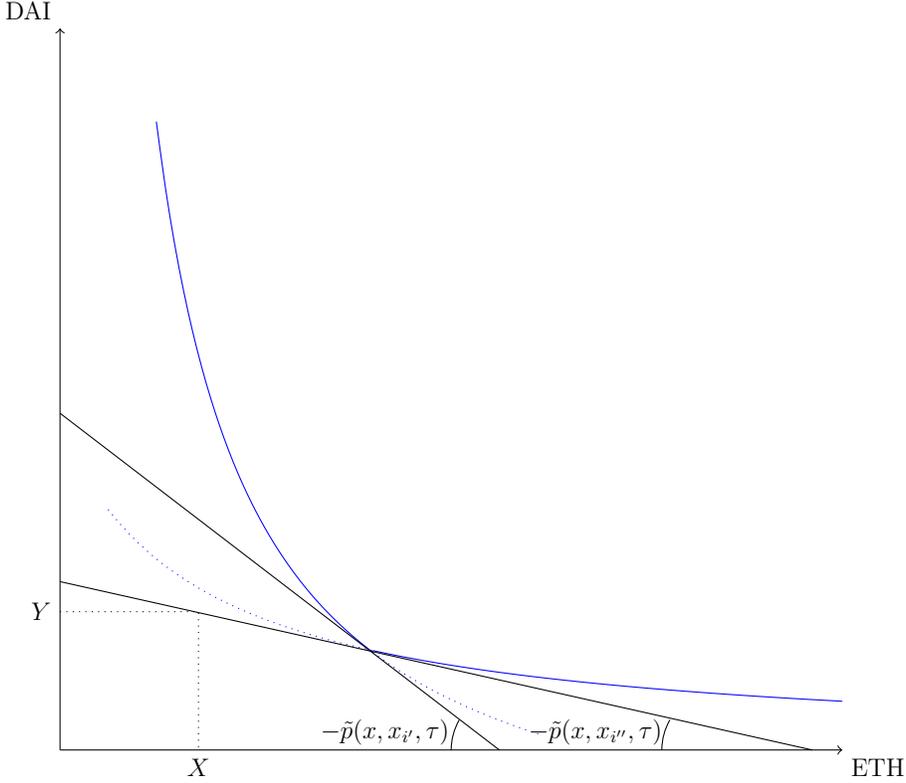
\begin{figure}[ht]
    \centering
    \begin{tikzpicture}[scale=1.6, every node/.style={scale=.8}]
   \def\t{.5}; 
   \def\QUSD{1.15}; 
   \def\QE{1.15}; 
   
    \draw[->] (0,0) -- (6.5,0) node[below right] {ETH};
    \draw[->] (0,0) -- (0,6) node[above left] {DAI};

 \draw[dotted] (0,\QUSD)node[left]{$Y$} -- (\QE,\QUSD) -- (\QE,0) node[below]{$X$};

      \draw[domain=.8:2.6,smooth,variable=\x, blue] plot ({\x},{5.1/\x-(\t/(1-\t))*\QUSD}) ; 
              \draw[dotted,domain=2.6:4,smooth,variable=\x, blue] plot ({\x},{5.1/\x-(\t/(1-\t))*\QUSD}) ; 
     
     \draw[domain=2.6:6.5,smooth,variable=\x, blue] plot ({\x},{(3.1/(\x+\QE*(\t/(1-\t)))}) ;
   \draw[dotted,domain=0.4:2.6,smooth,variable=\x, blue] plot ({\x},{(3.1/(\x+\QE*(\t/(1-\t)))}) ;

 \draw (0,1.4) --(6.25,0);

\draw(5,0) arc[start angle=0, end angle=-30, radius=-.5];
    \node at (4.45,0.15) {$-\tilde p(  x, x_{i''},\tau)$};



 \draw (0,2.8) --(3.65,0);

\draw(3.25,0) arc[start angle=0, end angle=-30, radius=-.5];
    \node at (2.7,0.15) {$-\tilde p(  x, x_{i'},\tau)$};

\end{tikzpicture}
    \caption{Effective price when there are both buy and sell orders, for $x<0$, $x_{i'}>0$, and $x_{i''}<0$.}
    \label{fig:function-max-with-fees-2}
\end{figure}

The key observation is that the price before fees depends on the trade occurring on the FM-AMM. We denote an individual order by $x_i$, and the net trade reaching the FM-AMM by $x$. Again, if $x<0$, then only $x(1-\tau)$ is exchanged on the FM-AMM, and the price before fee is $p^{FM-AMM}(x(1-\tau))$. If instead $x>0$, the full amount is exchanged on the FM-AMM and the price before fee is $p^{FM-AMM}(x)$. Given the price before fee, the sign of an order  $x_i$ determines its effective price, which is
\begin{equation}\label{eq: effective prices-other}
\tilde p(x, x_i, \tau) \equiv \begin{cases} 
\frac{1}{1-\tau} \cdot p^{FM-AMM}(  x) &\mbox{ if }  x\geq 0, x_i> 0 \\
(1-\tau) \cdot p^{FM-AMM}(  x) &\mbox{ if }  x\geq 0, x_i< 0 \\
\frac{1}{1-\tau} \cdot p^{FM-AMM}(  x(1-\tau)) &\mbox{ if }  x\leq 0, x_i > 0  \\
(1-\tau) \cdot p^{FM-AMM}(  x(1-\tau)) &\mbox{ if }  x\leq 0, x_i < 0\end{cases}
\end{equation}
See also Figure \ref{fig:function-max-with-fees-2}. It is easy to check that if $x_i$ and $x$ have the same sign, then the above expression is identical to \eqref{eq: effective prices}. The above equation is, therefore, a generalization of \eqref{eq: effective prices}. 
Also, some trades may be settled peer-to-peer even if $x=0$ and no trade is settled on the FM-AMM. The above expression implies that the price before fees when $x=0$ is $Y/X$, which is the same as in a standard CPAMM such as Uniswap.

Finally, note that for given $\tau$, the revenues earned by trades that are settled peer-to-peer is an exogenous transfer that does not affect the FM-AMM maximization of \eqref{eq: maximand with fees}. Hence, like in the previous case,  FM-AMM remains ``function maximizing'': for any given $p$ and $\tau$ it trades so to maximize  \eqref{eq: maximand with fees} and move ``up the curve'' (as in Figure  \ref{fig:function-max-with-fees}).

\section{The model}\label{sec: the model}

Equipped with the full description of an FM-AMM, we can now study its behavior in an environment with traders and arbitrageurs. We limit our analysis to the product function. 

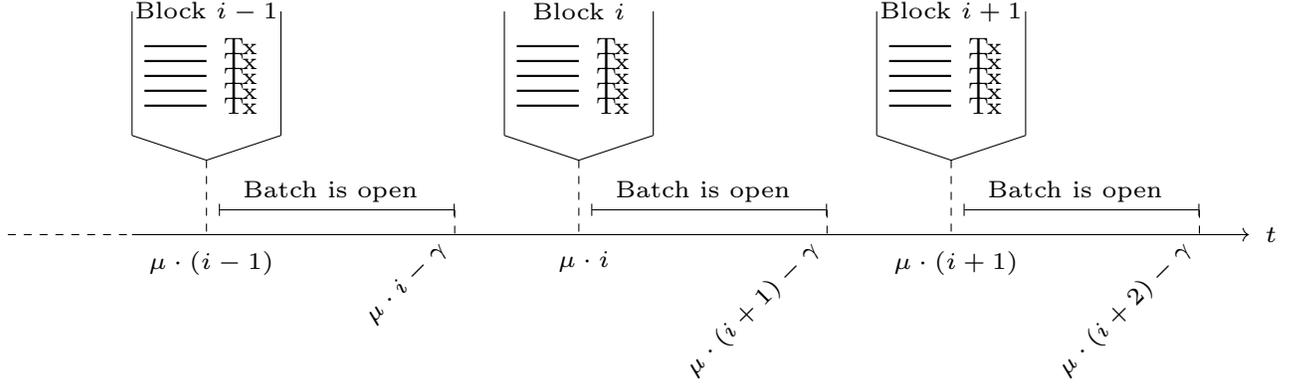
\begin{figure}
    \centering
   \begin{tikzpicture}[scale=1.65, every node/.style={scale=1.5}]
\usetikzlibrary {arrows.meta} 
\draw[->](0,0)--(9,0)node[right]{\tiny{$t$}};
\draw[dashed](-1,0)--(0,0);

\draw(0,.8+1)--(0,-.2+1);
\draw[thick](0.1,.4+1)--(.6,.4+1)node[right ]{\tiny{Tx}};
\draw[thick](0.1,.28+1)--(.6,.28+1)node[right ]{\tiny{Tx}};
\draw[thick](0.1,.16+1)--(.6,.16+1)node[right ]{\tiny{Tx}};
\draw[thick](0.1,.52+1)--(.6,.52+1)node[right ]{\tiny{Tx}};
\draw[thick](0.1,1.04)--(.6,1.04)node[right ]{\tiny{Tx}};
\draw(1.2,.8+1)--(1.2,-.2+1);
 \draw(0,-.2+1)--(.6,.6)--(1.2,-.2+1);
\draw[dashed](.6,.6)--(.6,0)node[below]{\tiny{ $\mu \cdot (i-1)$}};

 \draw [|-|] (.7,.2) -- (2.6,0.2);
\draw[dashed](2.6,0.2)--(2.6,0)node[left, rotate=45]{\tiny{$\mu \cdot i-\gamma$}};
\node at (1.6,.35) {\tiny{Batch is open}} ;

\node at (.6,1.8) {\tiny{Block $i-1$}} ;

\begin{scope}[xshift=3cm]
\draw(0,.8+1)--(0,-.2+1);
\draw[thick](0.1,.4+1)--(.6,.4+1)node[right ]{\tiny{Tx}};
\draw[thick](0.1,.28+1)--(.6,.28+1)node[right ]{\tiny{Tx}};
\draw[thick](0.1,.16+1)--(.6,.16+1)node[right ]{\tiny{Tx}};
\draw[thick](0.1,.52+1)--(.6,.52+1)node[right ]{\tiny{Tx}};
\draw[thick](0.1,1.04)--(.6,1.04)node[right ]{\tiny{Tx}};
\draw(1.2,.8+1)--(1.2,-.2+1);
 \draw(0,-.2+1)--(.6,.6)--(1.2,-.2+1);
\draw[dashed](.6,.6)--(.6,0)node[below]{\tiny{ $\mu \cdot i$}};
\node at (.6,1.8) {\tiny{Block $i$}} ;
 \draw [|-|] (.7,.2) -- (2.6,0.2);
\node at (1.6,.35) {\tiny{Batch is open}} ;
\draw[dashed](2.6,0.2)--(2.6,0)node[left, rotate=45]{\tiny{$\mu \cdot (i+1)-\gamma$}};
\end{scope}

\begin{scope}[xshift=6cm]
\draw(0,.8+1)--(0,-.2+1);
\draw[thick](0.1,.4+1)--(.6,.4+1)node[right ]{\tiny{Tx}};
\draw[thick](0.1,.28+1)--(.6,.28+1)node[right ]{\tiny{Tx}};
\draw[thick](0.1,.16+1)--(.6,.16+1)node[right ]{\tiny{Tx}};
\draw[thick](0.1,.52+1)--(.6,.52+1)node[right ]{\tiny{Tx}};
\draw[thick](0.1,1.04)--(.6,1.04)node[right ]{\tiny{Tx}};
\draw(1.2,.8+1)--(1.2,-.2+1);
 \draw(0,-.2+1)--(.6,.6)--(1.2,-.2+1);
\draw[dashed](.6,.6)--(.6,0)node[below]{\tiny{ $\mu \cdot (i+1)$}};
\node at (.6,1.8) {\tiny{Block $i+1$}} ;
 \draw [|-|] (.7,.2) -- (2.6,0.2);
\node at (1.6,.35) {\tiny{Batch is open}} ;
\draw[dashed](2.6,0.2)--(2.6,0)node[left, rotate=45]{\tiny{$\mu \cdot (i+2)-\gamma$}};

\end{scope}

\end{tikzpicture}
    \caption{Timeline}
    \label{fig:timeline}
\end{figure}

The timing is continuous. Every $\mu$ second, a new block is added to the blockchain. Traders can submit trades for inclusion in the batch anytime between the addition of a new block until $\gamma \in(0,\mu)$ seconds before the addition of the next block, where $\gamma$ is typically much smaller than $\mu$ (See Figure \ref{fig:timeline}). These trades are then settled in the next block. The batch settles trades peer-to-peer whenever possible, and the excess is settled on the FM-AMM. All trades on the same batch pay the same price before the fee and also the same fee (c.f., Section \ref{sec:fees}).   All trades submitted for inclusion in the batch are observable.

There are two types of traders: noise traders and arbitrageurs. In the batch preceding block $i \in\{1,2,...\}$,    each of $n$ noise traders submits a market order $a_j(i)$ for $j\in\{1,2,...,n\}$ to the FM-AMM, with the convention that if $a_j(i)>0$ then noise trader $j$ buys $a_j(i)$ ETH, while if $a_j(i)<0$ then noise trader $j$ sells $|a_j(i)|$ ETH.\footnote{The fact noise traders submit market orders also implies that each $\sum_j a_j(i)$ is small relative to the FM-AMM liquidity reserves. Otherwise, trading on the FM-AMM could have a large price impact, and we should treat orders from noise traders as limit orders. In this case, all our results continue to hold at the cost of additional notation.}  We also denote by $a_{+} (i) \equiv \sum_{j=1}^n \max\left\lbrace a_j(i), 0 \right\rbrace$ and $a_{-} (i)\equiv \sum_{j=1}^n \min\left\lbrace a_j(i), 0 \right\rbrace$ the total ETH demanded and supplied by noise traders, respectively, and by $a(i)=a_{+} (i)+a_{-} (i)$ their net demand. 

Besides noise traders, there are a large number of identical, cash-abundant, risk-neutral, competing arbitrageurs who can trade as part of the batch and on some external trading venue, assumed much larger and more liquid than the combination of noise traders and the FM-AMM. The equilibrium price for ETH on this external trading venue at time $t$ is $p^*(t)$ and is unaffected by trades on the FM-AMM. Arbitrageurs aim to profit from price differences between the FM-AMM and the external trading venues. Arbitrage opportunities will be intertemporal (over short intervals). Hence, for ease of derivations, we assume that arbitrageurs do not discount the future. Finally, the parameter $\gamma$ measures latency because it implies that arbitrageurs can use information up to $i \cdot \mu-\gamma$ when submitting trades on the batch that will settle on block $i$.\footnote{For ease of exposition, we assume that arbitrageurs have no latency, but the batch has latency $\gamma$. We could introduce a latency parameter for arbitrageurs and a different latency parameter for the batch, in which case $\gamma$ is the total latency.} We assume that the equilibrium price $p^*(t)$ is a continuous-time martingale. This assumption guarantees that the expectation of future prices is the current price; that is, for every $t'>t''$, we have $E_{t''}[p^*(t')]=p^*(t'')$ (where the expectation is taken with respect to the information available in period $t''$).

We are now ready to derive our main proposition. 

\begin{proposition}\label{prop: main}
Suppose that, at the end of block $i-1$, the reserves of the FM-AMM are $X$ and $Y$. Then, in the unique pure strategy equilibrium, in the subsequent batch:
\begin{itemize}
    \item If  $p^*(\mu \cdot i-\gamma)>\frac{1}{(1-\tau)}p^{FM-AMM}(a(i))$ or  $p^*(\mu \cdot i-\gamma)<(1-\tau)p^{FM-AMM}(a(i))$, then the arbitrageurs submit trade $r^*(i)$ such that
    \[
    \tilde p( a(i) +r^*(i), r^*(i), \tau)=p^*(\mu \cdot i-\gamma),
    \]
 In this case, we say that there is a rebalancing event.

    \item Otherwise, there is no rebalancing event, and arbitrageurs submit no trade, that is, $r^*(i)=0$. 

\end{itemize}
\end{proposition}

The formal proof of the proposition is in Appendix \ref{app:maths}. It is based on the observation that, because of competition between arbitrageurs, there cannot be any arbitrage opportunity left unexploited in equilibrium. If 
\[
p^*(\mu \cdot i-\gamma)>\frac{1}{(1-\tau)}p^{FM-AMM}(a(i)) \mbox{  or  } p^*(\mu \cdot i-\gamma)<(1-\tau)p^{FM-AMM}(a(i)),\]
then this intuition implies that arbitrageurs will rebalance the FM-AMM, that is, they will trade on FM-AMM  until its effective price (as faced by the arbitrageurs) is exactly  $p^*(\mu \cdot i-\gamma)$, which is also the expected price when the next block will be added to the blockchain. If instead 
\[
p^*(\mu \cdot i-\gamma)\in \left[(1-\tau)p^{FM-AMM}(a(i)), \frac{1}{(1-\tau)}p^{FM-AMM}(a(i))\right],
\]
then arbitrageurs cannot trade on the batch and make a profit because, given the trades submitted by noise traders, the buy price on the batch is higher than the equilibrium price, and the sell price on the batch is lower than the equilibrium price.

To conclude this section, we discuss how the FM-AMM performs in the presence of risk.  Consider period $\mu(i-1)$, when block $i-1$ is added to the blockchain. At that point in time, the future price $p^*(\mu \cdot i -\gamma)$ is a random variable with $E[p^*(\mu \cdot i -\gamma)]=p^*(\mu(i-1))$. 
In a traditional CFAMM, we know from the literature that arbitrage profits increase in the volatility of the price (see \citealp{capponi2021adoption}, \citealp{milionis2022automated}, and \citealp{milionis2023automated}). The intuition is that higher price volatility implies larger and more frequent trades by arbitrageurs.  Hence, in a CFAMM, the expected value of future LP holdings is lower when, from period $\mu \cdot (i-1)$ viewpoint, the variance of $p^*(\mu \cdot i -\gamma)$  is higher: a CFAMM is risk-averse in the value of its reserves.  However, the previous proposition shows that FM-AMM  trades with arbitrageurs at the expected equilibrium price. Hence, absent noise traders, FM-AMM is risk-neutral in the value of its reserves.\footnote{The presence of noise traders slightly complicates this logic: if there is a rebalancing event, noise traders who trade in the same direction as arbitrageurs also trade at the expected equilibrium price, while all other noise traders trade at a price that is a linear transformation of the equilibrium price.  If there is no rebalancing event, then the price at which noise traders trade is independent of the equilibrium price. The revenues earned from noise traders are, therefore, piecewise linear in the equilibrium price: they are linear if there is a rebalancing event, but they are flat if there is no rebalancing. They are neither globally concave nor globally convex, making it impossible to make a general statement about how revenues from noise traders are affected by the volatility of the equilibrium price.  }

At the same time, we discussed earlier how a CFAMM always trades to stay on the same function, while an FM-AMM trades to increase the value of its function. More precisely, for given prices and fee, FM-AMM trades to maximize the objective function  $U(  x,p, \tau)$, defined in \eqref{eq: maximand with fees}. The next proposition shows that this increase is larger the more risk there is. Hence, with respect to the value of the function, a CFAMM is risk-neutral (i.e. it always stays on the same function), while an FM-AMM is risk-loving.

\begin{proposition}[FM-AMM is risk loving]\label{prop: risk loving}
Consider two  probability distributions for the equilibrium price in period $\mu\cdot i -\gamma$, $F(p):R^+\rightarrow[0,1]$ and $G(p):R^+\rightarrow[0,1]$ having equal mean $p^*(\mu(i-1))$. Assume that  $F()$ is a mean-preserving spread of $G()$, that is, it is possible to write
\[
p^*_f = p^*_g + \epsilon
\]
where $p^*_f \sim F()$, $p^*_g \sim G()$ and $\epsilon$ is a shock with $E[\epsilon |p^*_g]=0$. Then, in expectation from period $\mu(i-1)$ viewpoint, the FM-AMM reaches a higher function in period $\mu\cdot i$ under distribution $F()$ than under distribution $G()$, that is 
\[
E_{F} [V( \tilde p( a(i)+r^*(i), \tau), \tau)|t=\mu(i-1)] \geq E_{G} [V(\tilde p( a(i)+r^*(i), \tau),\tau)|t=\mu(i-1)]
\]
where $V(p, \tau)=\max_{  x} U(  x,p, \tau)$. The inequality is strict if the probability of a rebalancing under distribution $F()$ is strictly positive.
\end{proposition}

See Appendix \ref{app:maths} for the proof. The proposition compares the expected value of the function under two distributions of the future price, where one distribution is a mean-preserving spread of the other. This ranking captures an intuitive notion of risk because one distribution can be derived from the other by adding some noise. If one distribution is a mean-preserving spread of another, the former distribution has a higher variance. Note that not all distributions can be ranked using mean-preserving spreads. However, it is usually the case that if both distributions belong to the same family (i.e., both normal), then ranking based on mean-preserving spreads coincides with the ranking based on variance.

\section{Empirical analysis}\label{sec: empirical analysis}

We complement our theoretical analysis by estimating the returns of providing liquidity to an FM-AMM. We do so by considering a counterfactual in which an FM-AMM existed during a specific period. We use Binance price data (together with our theoretical results) to simulate how arbitrageurs would have rebalanced our simulated FM-AMM. For the moment, we assume that there are no noise traders on the FM-AMM (that is $a_{-}(i)=a_+(i)=0$ for all $i$) and postpone to Section \ref{sec: empirical noise trading} the case of an FM-AMM that also receives noise traders. Hence, the estimated LP returns derived here should be considered the lower bound to the possible returns generated by an FM-AMM earning revenues from noise traders.  Also, we assume zero latency (that is, $\gamma=0$).\footnote{Introducing positive latency would not change our result but add noise to our estimation.}

We then compare the return of providing liquidity to our simulated FM-AMM to the empirical returns of providing liquidity to the corresponding Uniswap v3 pool. In this case, we assume that FM-AMM has the same fee as the corresponding Uniswap v3 pool and the batch frequency is 1 block. Because the simulated FM-AMM is not exploited by arbitrageurs but does not earn revenues from noise traders, this comparison establishes whether, on the Uniswap v3 pool we consider, arbitrageurs' profits exceed or fall short of the revenues generated by noise traders.
After precisely simulating an FM-AMM's returns without noise trading, we estimate the amount of noise trading volume (relative to Uniswap's volume) that an FM-AMM would need to receive to outperform Uniswap v3 on all pairs.
Finally, we study how the returns of FM-AMM LPs change with its fees.

\subsection{Details of the empirical analysis}

We select the top Uniswap v3 pools by trading volume from April 2023 to October 2023, excluding stablecoin pairs and tokens not traded on Binance, for a total of 11 pools. Among those, the highest-volume pairs are  WETH and WBTC exchanged with each other or with a stablecoin (7 pools), followed by less-traded tokens exchanged with WETH (4 pools).

For the same period and token pairs, we retrieve second-by-second price data from Binance. Note that on Uniswap, most tokens trade against either WETH or USDC. Instead, on Binance, almost all tokens trade against USDT, and only a few against USDC.\footnote{Furthermore, USDC was unavailable on Binance between September 2022 and March 2023 because, during this period, Binance would automatically convert all USDC deposits into its own stablecoin BUSD (see \url{https://www.binance.com/en/support/announcement/binance-to-auto-convert-usdc-usdp-tusd-to-busd-binance-usd-e62f703604a94538a1f1bc803b2d579f}). For this reason, we start our time series in April 2023.} As a consequence, for the most liquid token pairs we can use Binance prices directly, while for the less liquid token pairs we derive the Binance price of a token in ETH by combining the Binance prices of that token in USDT with the Binance price of ETH in USDT.
The implicit assumption is that arbitrageurs can trade on multiple Binance markets.  Proposition \ref{prop: main} then allows us to compute the size and direction of the trade that would have rebalanced an FM-AMM to the Binance price. These rebalancing trades determine the evolution of the FM-AMM reserves and the return of its liquidity providers.

Then, for each of our selected Uniswap v3 pools, we calculate the return of a simulated \textit{unconcentrated} liquidity position (i.e.,  a position over the entire price range $[0, \infty]$). Note that on Uniswap v3, a liquidity position can be \textit{concentrated}, meaning that the liquidity is available only over a certain price range. Also, when a swap occurs, the fees collected are distributed proportionally among the liquidity available around the price at which the swap occurred.
Therefore, to compute the fees earned by our simulated liquidity position, we retrieve all swap transactions on the pool in question from an Ethereum node. For each of these swaps, we collect data on the fee paid and the liquidity available around the price just after the swap.
We can then calculate the fees our simulated liquidity position would have earned.

Our method is based on three assumptions. First, the simulated liquidity position is too small to affect traders' incentives to trade and other LPs' incentives to provide liquidity. 
Second, we implicitly assume that the price stays within the same ticks during a swap transaction. This is inaccurate because when a swap causes a large price movement,  
a part of the fee is shared among the liquidity available around the \textit{initial} price (and not only around the final price). 
If the liquidity available at the initial price is less (more) than that available at the final price, then our method underestimates (overestimates) the fees earned by our liquidity position, leading to some non-systematic inaccuracies in our estimation.\footnote{Besides being non-systematic, the inaccuracies introduced are likely to be minimal. To illustrate this, we calculate that the difference between assuming liquidity to be constant over a full block instead of over each swap is 0.01\% over 6 months for the WETH-USDC 0.05\% pool. We expect the inaccuracies from assuming the liquidity to be constant during each swap to be even smaller.} Third, we assume that the fees earned by our simulated liquidity position are automatically and costlessly contributed to the pool as additional liquidity. However, on Uniswap v3, fees accrue outside the AMM. Using them as additional liquidity is possible by exchanging a fraction of the fees earned (in a single token) in the other tokens and then contributing them as liquidity. This generates gas costs and additional AMM fees, which we ignore in our calculation. Our method, therefore, overestimates the return of providing liquidity on Uniswap v3


Finally, our results do not depend on the size of the initial liquidity position. On Uniswap v3, a larger initial position earns proportionally more fees, but its ROI is the same. Similarly, on an FM-AMM, the size of the rebalancing trade scales proportionally with the available liquidity so that, again, its ROI is independent of its initial size. Also, as already discussed, for both Uniswap v3 and the FM-AMM, we consider a non-concentrated liquidity position. If both positions are concentrated in the same (symmetrical) way, both Uniswap v3 fees and FM-AMM returns increase by the same factor as long as the price does not go out of range. 
So the comparison does not change, and the full-range comparison already constitutes a general comparison.

\subsection{Results}

\paragraph{Arbitrageurs' profits vs.\ Uniswap fees}

\begin{figure}[htp]
    \centering
    \begin{subfigure}{\textwidth}
        \includegraphics[width=\textwidth]{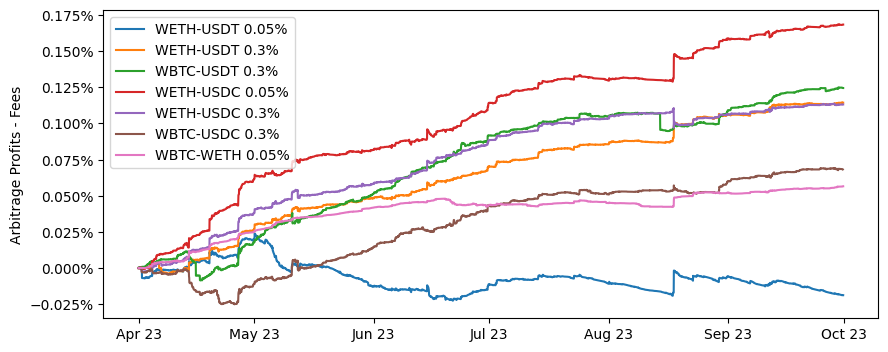}
        \caption{Highly traded pairs between ETH, BTC and stablecoins.}
        \label{fig:lvr_fee_comp_majors}
    \end{subfigure}%
    
    \begin{subfigure}{\textwidth}
        \includegraphics[width=\textwidth]{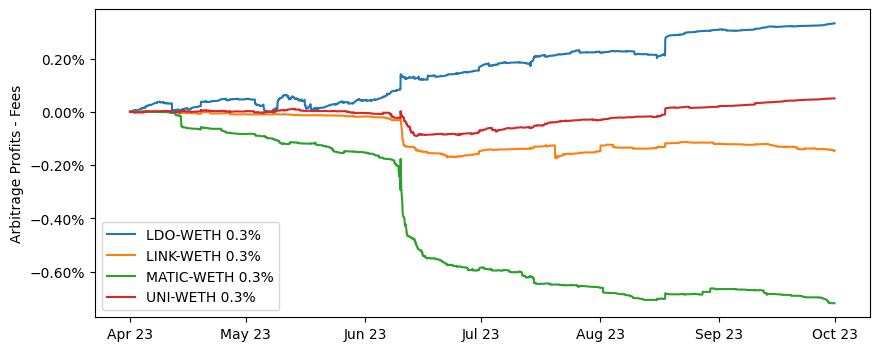}
        \caption{Less traded pairs, with Binance prices obtained by combining the prices of the two respective USDT pairs.}
        \label{fig:lvr_fee_comp_altcoins}
    \end{subfigure}
    
    \caption{Comparison over six months (April 2023 to October 2023): The difference between the return of providing liquidity to an FM-AMM and on Uniswap v3. Note that, relative to FM-AMM LPs, Uniswap v3 LPs lose to arbitrageurs but gain fees from noise traders. Comparing the return of the two liquidity positions is, therefore, equivalent to arbitrageurs' profits minus fees collected from noise traders on Uniswap v3. }
    \label{fig:lvr_fee_comp}
\end{figure}

Figure \ref{fig:lvr_fee_comp} shows the difference in cumulative returns between a simulated FM-AMM and the corresponding Uniswap v3 pool. For each trading pair, the FM-AMM trading fee is set equal to the fee of the Uniswap v3 pool we are comparing to.

For most high-volume pairs, providing liquidity on an FM-AMM consistently outperforms providing liquidity on Uniswap (see Figure \ref{fig:lvr_fee_comp_majors}). The ETH-USDT 0.05\% pool is the only exception, as the FM-AMM and Uniswap v3 returns are approximately equal. Remember that the comparison between FM-AMM and Uniswap v3 illustrates whether trading fees exceed or fall short of arbitrage profits on Uniswap V3. With this respect, our results indicate that providing liquidity to the largest and most traded Uniswap v3 pools is unprofitable: trading fees in these pools do not sufficiently compensate liquidity providers for arbitrage losses. On the other hand, for pools containing less-traded tokens, the results are mixed, with Uniswap v3 outperforming FM-AMM on some token pairs (see Figure \ref{fig:lvr_fee_comp_altcoins}). However, the absolute difference between cumulative returns is generally quite small. 
For all pools, the FM-AMM receiving 30\% of Uniswap v3's total trading volume as noise trading volume would suffice for the FM-AMM to outperform. This scenario seems realistic, as the share of noise trading on Uniswap is likely higher than 30\%\footnote{This is based on \cite{heimbach2024non} finding that 30\% of Uniswap volume can be attributed to non-atomic arbitrage, while less than 3\% each are sandwich attack and atomic arbitrage volume according to labeling by \url{https://zeromev.org/}.}.

\paragraph{FM-AMM fees}
So far, we assumed that the simulated FM-AMM has the same fee as the corresponding Uniswap pool. We now explore whether choosing a different fee would have generated higher returns for FM-AMM LPs.


\begin{figure}[htp]
  \centering

  \begin{subfigure}[b]{0.49\textwidth}
    \includegraphics[width=\textwidth]{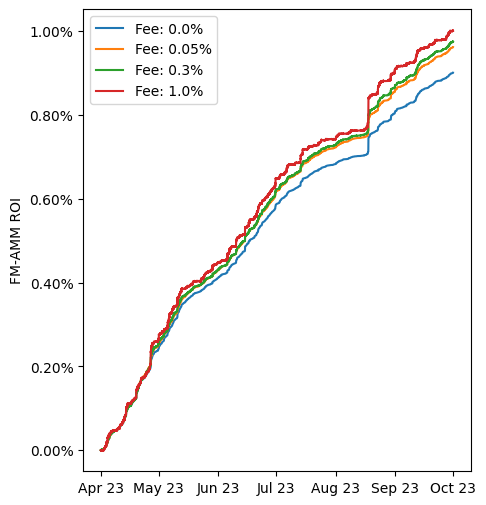}
    \caption{ETH-USDT}
    \label{fig:fees_ETH_USDT}
  \end{subfigure}
  \hfill
  \begin{subfigure}[b]{0.49\textwidth}
    \includegraphics[width=\textwidth]{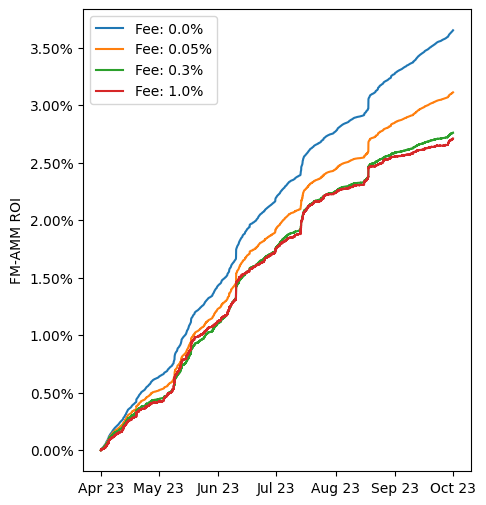}
    \caption{LDO-ETH}
    \label{fig:fees_LDO_ETH}
  \end{subfigure}

  \caption{Return on providing liquidity to an FM-AMM, for different fees: 0.0\%, 0.05\%, 0.3\%, 1.0\%}
  \label{fig:LVR_fees}
\end{figure}

Remember that the fee affects the size of the rebalancing trade, with higher fees implying smaller rebalancing trades (cf equation \ref{eq: effective prices}), which always occur at the new equilibrium price.  The fee also determines whether arbitrageurs rebalance the pool, with higher fees implying a lower probability that the pool will be rebalanced. In turn, the probability of rebalancing impacts the return of providing liquidity to FM-AMM because of path dependence. For example, an FM-AMM earns more when it settles one large batch instead of two smaller batches trading in the same direction. In this case,  rebalancing less frequently may be beneficial. However, settling a single large batch in which opposite trades net out may generate little or no trade (and hence little or no benefit to the FM-AMM), while settling two smaller batches trading in opposite directions moves the FM-AMM ``up the curve'' each time. In this case, rebalancing more frequently may be more beneficial. 

Figure \ref{fig:LVR_fees} shows two examples, one in which a fee of zero is optimal and another in which a strictly positive fee is optimal (see Appendix \ref{app:figures} for additional token pairs).  In most cases, the optimal fee is zero, which implies more frequent rebalancing. There are exceptions, though, where LP returns on an FM-AMM would have been highest for a larger-than-zero fee. 

Of course, these results do not necessarily extend to an FM-AMM with a positive volume of noise trading. They, however, illustrate that in the previous comparison with Uniswap v3, it is restrictive to use for FM-AMM the same fee as for Uniswap v3. For example, the MATIC-ETH FM-AMM with no noise traders and the same fee as Uniswap v3 underperforms Uniswap v3 by 0.65\%, which is reduced to 0.39\% if the FM-AMM has zero fee.

\subsection{Extension: FM-AMM with noise trading}\label{sec: empirical noise trading}

One important limitation of our data is that we do not know which Uniswap transactions are noise trading, arbitrageurs rebalancing the pool, or sandwich attacks. However, \cite{heimbach2024non} identifies 30\% of the total volume on Unswap v3 as non-atomic arbitrage. Also, according to labeling by \url{https://zeromev.org/}, sandwich attack and atomic arbitrage volume correspond to 6\% of Unswap v3 volume. If we extrapolate from Unsiwap v3 in general to our study pools, then in each of our study pools, approximately 60\% of trading volume is noise trading.

\begin{table}[ht]
    \centering
    \begin{tabular}{|c| c c c c|} 
        \hline
        FM-AMM noise trading volume & none & 10\% (16\%) & 30\% (50\%) & 50\% (83\%) \\ \hline
        \hline
            WETH-USDT 0.05\%& -0.01\%& 0.09\%& 0.28\%& 0.48\%\\ \hline
            WETH-USDT 0.3\%& 0.12\%& 0.20\%& 0.37\%& 0.55\%\\ \hline
            WBTC-USDT 0.3\%& 0.14\%& 0.21\%& 0.36\%& 0.50\%\\ \hline
            WETH-USDC 0.05\%& 0.18\%& 0.26\%& 0.42\%& 0.58\%\\ \hline
            WETH-USDC 0.3\%& 0.12\%& 0.20\%& 0.37\%& 0.55\%\\ \hline
            WBTC-USDC 0.3\%& 0.07\%& 0.15\%& 0.31\%& 0.47\%\\ \hline
            WBTC-WETH 0.05\%& 0.06\%& 0.08\%& 0.11\%& 0.15\%\\ \hline
            LDO-WETH 0.3\%& 0.38\%& 0.62\%& 1.10\%& 1.57\%\\ \hline
            LINK-WETH 0.3\%& -0.08\%& 0.06\%& 0.32\%& 0.59\%\\ \hline
            MATIC-WETH 0.3\%& -0.65\%& -0.44\%& -0.01\%& 0.41\%\\ \hline
            UNI-WETH 0.3\%& 0.08\%& 0.19\%& 0.40\%& 0.61\%\\ \hline
    \end{tabular}
    \caption{Difference between FM-AMM and Uniswap v3 returns for a full-range liquidity position over 6 months (April 2023 - October 2023) for different scenarios of how much noise trading volume the FM-AMM receives as a proportion of total Uniswap trade volume (approximate noise trading volume). Positive values indicate a higher return on the FM-AMM.}
    \label{tbl:fm_amm_uniswap}
\end{table}

With this approximation in mind, we can repeat our empirical analysis under different assumptions regarding FM-AMM noise trading volume relative to the corresponding Uniswap v3 pool's noise trading volume. We report the results in Table \ref{tbl:fm_amm_uniswap}. In all our study pools, the return of providing liquidity on an FM-AMM exceeds that of providing liquidity on Uniswap as long as FM-AMM's noise trading volume exceeds 30\% of the volume of the corresponding Uniswap pool, which under our approximation corresponds to 50\% of that Uniswap pool's noise trading volume.

\section{Discussion: attack model.}\label{sec: attack}

In our model, we assumed that batching is an off-chain component of the AMM. Although other implementations are possible (see Section \ref{sec: batching}), it is worth discussing a  vulnerability introduced by this assumption: the malicious ``batch operator''. 

Because the batching process is done off-chain, an entity must collect trades, settle them peer-to-peer whenever possible, and settle the rest on the FM-AMM as a single transaction. We call this entity the ``batch operator''. The smart contract managing the FM-AMM can guarantee that the batch operator cannot perform multiple transactions per block on the FM-AMM (hence preventing manipulation by splitting trades as described in Section \ref{sec: path dependence}). However, the batch operator has full discretion over what transactions are included in the batch and could exploit the FM-AMM by suppressing trades and acting as the ``unique arbitrageur.'' Similarly to the first arbitrageur reaching a CFAMM after a change in the equilibrium price, the batch operator is the only entity that can rebalance the pool and will exploit this opportunity to maximize profits. 

More precisely, suppose the external price is $p^*$ and that the liquidity reserves on the FM-AMM are $X$ and $Y$ (in ETH and DAI, respectively).  The malicious batch operator trades $x$ so to maximize its profits:
\[
\max_x \left\lbrace x\cdot p^* - x \cdot \frac{Y}{X-2x}   \right\rbrace = \frac{Y+p^* X}{2} -\sqrt{XYp^*}.
\]
Compare this to the profits earned by an arbitrageur rebalancing a CPAMM also with reserves $X$ and $Y$, which are
\[
\max_x \left\lbrace x\cdot p^* - x \cdot \frac{Y}{X-x}   \right\rbrace = Y+p^* X - 2\sqrt{XYp^*}.
\]
Note that arbitrageurs' and batch operators' profits are at the expense of their respective LPs. The above derivations, therefore, show that the loss that FM-AMM's LPs may face from a malicious batch operator are half of the losses that CPAMM's LPs routinely face from arbitrageurs. 

To conclude, note that the batch operator's behavior is observable ex-post on-chain. It is, therefore, possible to introduce punishments on the batch operator in case it misbehaves. There could be social punishments such as LPs withdrawing their liquidity (and hence making FM-AMM useless), automated slashing of dedicated funds, or even legal remedies if the batch operator is a legal entity.

 \section{Conclusion}

This paper studies the design of AMM when trades are batched before reaching the AMM. The key observation is that, because of batching, the AMM does not need to satisfy path independence, which implies that the design space is larger than without batching. In particular, it is possible to design a function maximizing AMM (FM-AMM), which, for given prices,  always trades to be on the highest possible value of a given function. At the same time, batching creates competition between informed arbitrageurs. As a result of this competition, the price at which the FM-AMM trades is always equal to the equilibrium price (assumed determined on some very liquid location and exogenous to the FM-AMM). Hence, whereas in a traditional CFAMM, arbitrageurs earn profits at the expense of liquidity providers, in an FM-AMM these profits remain with its LPs. At the same time, sandwich attacks are eliminated because all trades within the same batch occur at the same price, equal to the exogenously determined equilibrium price.  We use Binance price data for 11 token pairs over a period of six months to simulate a lower bound to the return of providing liquidity to an FM-AMM. We find that this lower bound generally outperforms providing the same liquidity on Uniswap v3, and that the absolute differences between the two are small.

Our results are robust to several extensions. In particular, in the companion paper \cite{canidio_et_al:LIPIcs.AFT.2023.24}, we study the case in which the batch operator can settle the trades on the batch both on the FM-AMM and also on some other CFAMM (while the FM-AMM can only be accessed via the batch). Also in that version of the model, competition among arbitrageurs guarantees that the FM-AMM always trades at the equilibrium price. The only difference is that arbitrageurs may rebalance the FM-AMM and the CFAMM by trading with the batch and then simultaneously backrunning the batch on the CFAMM. In that case,  arbitrageurs earn strictly positive profits at the expense of the CFAMM liquidity providers.


As discussed in the introduction, by eliminating sandwich attacks and arbitrage profits, an FM-AMM also eliminates the vast majority of MEV. If it were widely adopted, the only remaining  MEV source of measurable size would be liquidation in the context of lending protocols. In blockchain-based lending protocols, users borrow a given token by pledging a different token as collateral. The collateral is liquidated when its value drops below a threshold: the protocol sells it to whoever pays the liquidation price---usually well below its market value---and returns the borrowed amount. This naturally creates competition between liquidation bots and, therefore, MEV. We conjecture that if the liquidation process were forced through the FM-AMM batch, the liquidation would occur at the fair equilibrium price, eliminating this remaining source of MEV. Fully exploring this possibility is left for future work.

\appendix

\section{Mathematical derivations}\label{app:maths}

\begin{proof}[Proof of Proposition \ref{prop: main}]

Suppose nose traders collectively submit trade $a(i)>0$ (the case $a(i)<0$ is analogous). If arbitrageurs also submit a trade $r(i)>0$, then by \eqref{eq: effective prices-other} the effective price they pay is $\frac{Y}{(1-\tau)(X-2(a(i)+r(i)))}$. If instead they submit $r(i)<0$ with $a(i)+r(i)\leq 0$, then the effective price they pay is $\frac{Y(1-\tau)}{X-2(1-\tau)(a(i)+r(i))}$. If instead they submit $r(i)<0$ with $a(i)+r(i)\geq0$, then the effective price they pay is $\frac{Y(1-\tau)}{X-2(a(i)+r(i))}$. The price faced by arbitrageurs as a function of their trade is therefore discontinuous at $r(i)=0$, where arbitrageurs shift from paying the buy price $\frac{Y}{(1-\tau)(X-2a(i))}$ to paying the sell price $\frac{Y(1-\tau)}{X-2a(i)}$. For $r(i) \neq 0$, instead, the effective price paid by the arbitrageurs is continuous, strictly increasing, goes to zero for $r(i)\rightarrow -\infty$ and to infinity for $r(i) \rightarrow X-a(i)$. 

Hence, if  $p^*(\mu \cdot i-\gamma)>\frac{Y}{(1-\tau)(X-2a(i))}$ or $p^*(\mu \cdot i-\gamma)<\frac{Y(1-\tau)}{X-2a(i)}$, then there exists a unique $r^*(i)$ such that
    \[
    \tilde p( a(i) +r^*(i), r^*(i), \tau)=p^*(\mu \cdot i-\gamma),
    \]
In this case,  $r^*(i)$  is an equilibrium because no arbitrageur can be better off by trading more or less with the batch. The fact that such an equilibrium is unique can be established by contradiction: suppose the equilibrium is $r'$ with $\tilde p( a(i) +r', r', \tau)\neq p^*(k\cdot i -\gamma)$. Then by the fact that $\tilde p( a(i) +r', r' \tau)$  is locally continuous in $r'$, an arbitrageur could submit an additional trade $k$ such that   $\tilde p( a(i) +r' +k,r' +k, \tau) \neq p^*(k\cdot i -\gamma)$ and earn strictly positive profits, which implies that $r'$ is not an equilibrium. 

Finally, it is easy to check that if  $p^*(k\cdot i -\gamma)\in [\frac{Y}{(1-\tau)(X-2 a(i))}, \frac{(1-\tau) Y}{X-2 a(i)}]$, then arbitrageurs have no profitable trade opportunity: if arbitrageurs also buy ETH they pay at least $\frac{(1-\tau) Y}{X-2 a(i)}$, which is greater than the equilibrium price; if arbitrageurs sell ETH they receive at most  $\frac{Y}{(1-\tau)(X-2 a(i))}$ per ETH sold, which is less than the equilibrium price.

\end{proof}

\begin{proof}[Proof of Proposition \ref{prop: risk loving}]

Note that the volume of noise trading is independent of the price. Hence, for given prices, a FM-AMM maximizes $V(p, \tau)=\max_{  x} U(  x+a(i),p, \tau)$ with demand function (net of noise trading) given by $x(p,\tau)\equiv \mbox{argmax}_{  x} U(  x+a(i),p, \tau) \neq 0$.

The first observation is that the value function $V(p, \tau)=\max_{  x} U(  x+a(i),p, \tau)$ is convex in $p$, strictly so if there is strictly positive trade (i.e., if $  x(p, \tau) \neq 0$). To see this, use the envelope theorem to write:
\[
\frac{\partial V(p, \tau)}{\partial p} = \begin{cases} 
 \left(X -   x(p, \tau)-a(i)\right)    x(p, \tau)         &\mbox{ if }   x(p, \tau)>0 \\
 \left(\frac{X}{1-\tau} -  x(p, \tau)-a(i)\right)    x(p, \tau)  &\mbox{ if }   x(p, \tau)<0 \\
 0 &\mbox{ if }   x(p, \tau)=0 
 \end{cases}
\]
so that 
\[
\frac{\partial^2 V(p, \tau)}{\partial p^2} = \begin{cases} 
\frac{\partial   x(p,\tau) }{\partial p} \left(X -a(i) - 2   x(p, \tau)\right)          &\mbox{ if }   x(p, \tau)>0 \\
 \frac{\partial   x(p,\tau) }{\partial p} \left(\frac{X}{1-\tau} -a(i) - 2   x(p, \tau)\right)   &\mbox{ if }   x(p, \tau)<0 \\
 0 &\mbox{ if }   x(p, \tau)=0 
 \end{cases}
\]
Because $p$ and $  x$ are strict complements in the objective function, by Topkis's theorem, $\frac{\partial   x(p,\tau) }{\partial p}>0$ whenever $  x(p, \tau) \neq 0$. Also, the FM-AMM always trades so that $\left(X-a(i) - 2   x(p, \tau)\right) >0$ and $\left(\frac{X}{1-\tau} - a(i) - 2   x(p, \tau)\right)>0$. It follows that $V(p,\tau)$ is strictly convex in $p$ whenever $  x(p, \tau) \neq 0$. 

Next, note that if there is a rebalancing event, then $p=p^*(\mu\cdot i -\gamma)$. If there is no rebalancing event, then $V(p, \tau)$ is independent of the price. Because  $F()$ is a mean-preserving spread of $G()$, then
\[
E_F [V(p, \tau)] \geq E_G [V(p,\tau)]
\]
with strict inequality as long as $  x(p,\tau)>0$ (i.e., there is a rebalancing event) for some realization of $p^*(\mu\cdot i -\gamma)$.
\end{proof}

\section{Extra figures}\label{app:figures}

\begin{figure}[htp]
  \centering

  \begin{subfigure}[b]{0.45\textwidth}
    \includegraphics[width=\textwidth]{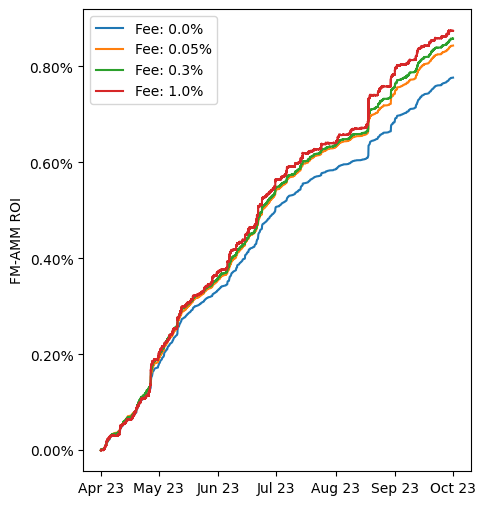}
    \caption{BTC-USDT}
    \label{fig:fees_BTC_USDT}
  \end{subfigure}
  \hfill
  \begin{subfigure}[b]{0.45\textwidth}
    \includegraphics[width=\textwidth]{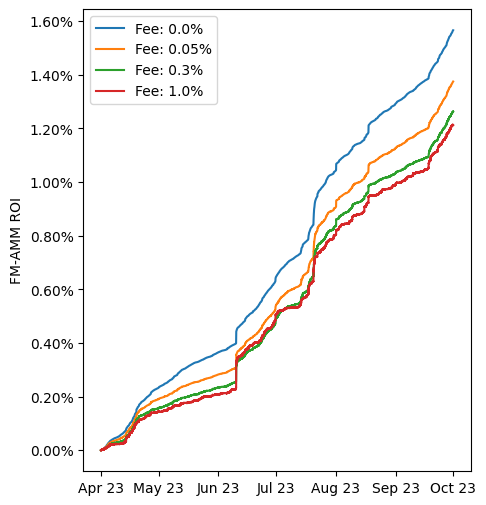}
    \caption{LINK-ETH}
    \label{fig:fees_LINK_ETH}
  \end{subfigure}

  \begin{subfigure}[b]{0.45\textwidth}
    \includegraphics[width=\textwidth]{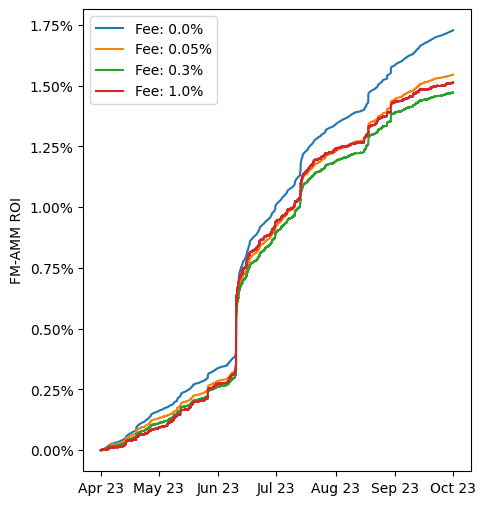}
    \caption{MATIC-ETH}
    \label{fig:fees_MATIC_ETH}
  \end{subfigure}
  \hfill
  \begin{subfigure}[b]{0.45\textwidth}
    \includegraphics[width=\textwidth]{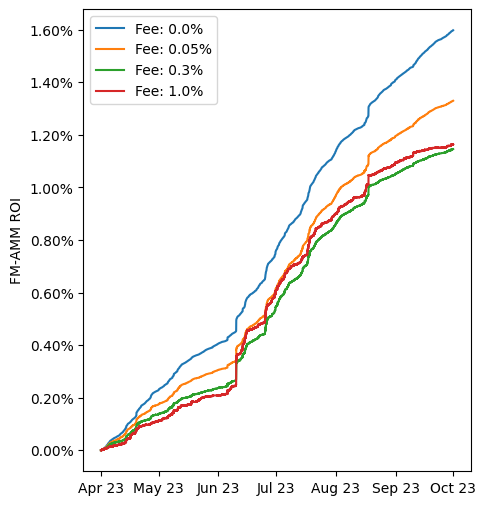}
    \caption{UNI-ETH}
    \label{fig:fees_UNI_ETH}
  \end{subfigure}

  \caption{Return on providing liquidity to an FM-AMM, for different fees: 0.0\%, 0.05\%, 0.3\%, 1.0\%}
  \label{fig:LVR_fees_extra}
\end{figure}

 \newpage

\bibliography{bib}{}

\begin{thebibliography}{}

\bibitem[\protect\citeauthoryear{Aoyagi}{Aoyagi}{2020}]{aoyagi2020liquidity}
Aoyagi, J. (2020).
\newblock Liquidity provision by automated market makers.
\newblock {\em working paper\/}.

\bibitem[\protect\citeauthoryear{Aquilina, Budish, and O'neill}{Aquilina
  et~al.}{2022}]{aquilina2022quantifying}
Aquilina, M., E.~Budish, and P.~O'neill (2022).
\newblock Quantifying the high-frequency trading ``arms race''.
\newblock {\em The Quarterly Journal of Economics\/}~{\em 137\/}(1), 493--564.

\bibitem[\protect\citeauthoryear{Breidenbach, Daian, Tram{\`e}r, and
  Juels}{Breidenbach et~al.}{2018}]{breidenbach2018enter}
Breidenbach, L., P.~Daian, F.~Tram{\`e}r, and A.~Juels (2018).
\newblock Enter the hydra: Towards principled bug bounties and
  $\{$Exploit-Resistant$\}$ smart contracts.
\newblock In {\em 27th USENIX Security Symposium (USENIX Security 18)}, pp.\
  1335--1352.

\bibitem[\protect\citeauthoryear{Budish, Cramton, and Shim}{Budish
  et~al.}{2015}]{budish2015high}
Budish, E., P.~Cramton, and J.~Shim (2015).
\newblock The high-frequency trading arms race: Frequent batch auctions as a
  market design response.
\newblock {\em The Quarterly Journal of Economics\/}~{\em 130\/}(4),
  1547--1621.

\bibitem[\protect\citeauthoryear{Canidio and Danos}{Canidio and
  Danos}{ming}]{canidio2023commitment}
Canidio, A. and V.~Danos (forthcoming).
\newblock Commitment against front-running attacks.
\newblock {\em Management Science\/}.

\bibitem[\protect\citeauthoryear{Canidio and Fritsch}{Canidio and
  Fritsch}{2023}]{canidio_et_al:LIPIcs.AFT.2023.24}
Canidio, A. and R.~Fritsch (2023).
\newblock {Batching Trades on Automated Market Makers}.
\newblock In J.~Bonneau and S.~M. Weinberg (Eds.), {\em 5th Conference on
  Advances in Financial Technologies (AFT 2023)}, Volume 282 of {\em Leibniz
  International Proceedings in Informatics (LIPIcs)}, Dagstuhl, Germany, pp.\
  24:1--24:17. Schloss Dagstuhl -- Leibniz-Zentrum f{\"u}r Informatik.

\bibitem[\protect\citeauthoryear{Capponi and Jia}{Capponi and
  Jia}{2021}]{capponi2021adoption}
Capponi, A. and R.~Jia (2021).
\newblock The adoption of blockchain-based decentralized exchanges.
\newblock {\em arXiv preprint arXiv:2103.08842\/}.

\bibitem[\protect\citeauthoryear{Chainlink}{Chainlink}{2020}]{chainlink}
Chainlink (2020).
\newblock What is the blockchain oracle problem?
\newblock Retrieved from \url{https://chain.link/education-hub/oracle-problem}
  on May 24, 2023.
\newblock Online forum post.

\bibitem[\protect\citeauthoryear{Della~Penna}{Della~Penna}{2022}]{dellapenna2022mev}
Della~Penna, N. (2022, September 1).
\newblock Mev minimizing amm (minmev amm).
\newblock Retrieved from
  \url{https://ethresear.ch/t/mev-minimizing-amm-minmev-amm/13775} on May 24,
  2023.
\newblock Online forum post.

\bibitem[\protect\citeauthoryear{Ferreira and Parkes}{Ferreira and
  Parkes}{2023}]{ferreira2023credible}
Ferreira, M. V.~X. and D.~C. Parkes (2023).
\newblock Credible decentralized exchange design via verifiable sequencing
  rules.

\bibitem[\protect\citeauthoryear{Foley, O'Neill, and Putnins}{Foley
  et~al.}{2022}]{foley2022can}
Foley, S., P.~O'Neill, and T.~Putnins (2022).
\newblock Can markets be fully automated? evidence from an automated market
  maker.
\newblock Technical report, Working Paper, Macquarie University.

\bibitem[\protect\citeauthoryear{Forgy and Lau}{Forgy and
  Lau}{2021}]{forgy2021family}
Forgy, E. and L.~Lau (2021).
\newblock A family of multi-asset automated market makers.
\newblock {\em arXiv preprint arXiv:2111.08115\/}.

\bibitem[\protect\citeauthoryear{Foucault}{Foucault}{1999}]{foucault1999order}
Foucault, T. (1999).
\newblock Order flow composition and trading costs in a dynamic limit order
  market.
\newblock {\em Journal of Financial markets\/}~{\em 2\/}(2), 99--134.

\bibitem[\protect\citeauthoryear{Gans and Holden}{Gans and
  Holden}{2022}]{gans2022solomonic}
Gans, J.~S. and R.~T. Holden (2022).
\newblock A solomonic solution to ownership disputes: An application to
  blockchain front-running.
\newblock Technical report, National Bureau of Economic Research.

\bibitem[\protect\citeauthoryear{Heimbach, Kiffer, Ferreira~Torres, and
  Wattenhofer}{Heimbach et~al.}{2023}]{heimbach2023ethereum}
Heimbach, L., L.~Kiffer, C.~Ferreira~Torres, and R.~Wattenhofer (2023).
\newblock Ethereum's proposer-builder separation: Promises and realities.
\newblock In {\em Proceedings of the 2023 ACM on Internet Measurement
  Conference}, pp.\  406--420.

\bibitem[\protect\citeauthoryear{Heimbach, Pahari, and Schertenleib}{Heimbach
  et~al.}{2024}]{heimbach2024non}
Heimbach, L., V.~Pahari, and E.~Schertenleib (2024).
\newblock Non-atomic arbitrage in decentralized finance.
\newblock {\em arXiv preprint arXiv:2401.01622\/}.

\bibitem[\protect\citeauthoryear{Heimbach, Schertenleib, and
  Wattenhofer}{Heimbach et~al.}{2022}]{heimbach2022risks}
Heimbach, L., E.~Schertenleib, and R.~Wattenhofer (2022).
\newblock Risks and returns of uniswap v3 liquidity providers.
\newblock {\em arXiv preprint arXiv:2205.08904\/}.

\bibitem[\protect\citeauthoryear{Heimbach, Wang, and Wattenhofer}{Heimbach
  et~al.}{2021}]{heimbach2021behavior}
Heimbach, L., Y.~Wang, and R.~Wattenhofer (2021).
\newblock Behavior of liquidity providers in decentralized exchanges.

\bibitem[\protect\citeauthoryear{Johnson, Diamandis, Evans, de~Valence, and
  Angeris}{Johnson et~al.}{2023}]{johnson2023concave}
Johnson, N.~A., T.~Diamandis, A.~Evans, H.~de~Valence, and G.~Angeris (2023).
\newblock Concave pro-rata games.
\newblock {\em arXiv preprint arXiv:2302.02126\/}.

\bibitem[\protect\citeauthoryear{Josojo}{Josojo}{2022}]{josojo2022mev}
Josojo (2022, August 4).
\newblock Mev capturing amm (mcamm).
\newblock Retrieved from
  \url{https://ethresear.ch/t/mev-capturing-amm-mcamm/13336} on May 24, 2023.
\newblock Online forum post.

\bibitem[\protect\citeauthoryear{Lehar and Parlour}{Lehar and
  Parlour}{2021}]{lehar2021decentralized}
Lehar, A. and C.~A. Parlour (2021).
\newblock Decentralized exchanges.
\newblock {\em working paper\/}.

\bibitem[\protect\citeauthoryear{Leupold}{Leupold}{2022}]{leupold2022cow}
Leupold, F. (2022, November 1).
\newblock Cow native amms (aka surplus capturing amms with single price
  clearing).
\newblock Retrieved from
  \url{https://forum.cow.fi/t/cow-native-amms-aka-surplus-capturing-amms-with-single-price-clearing/1219/1}
  on May 24, 2023.
\newblock Online forum post.

\bibitem[\protect\citeauthoryear{Loesch, Hindman, Richardson, and Welch}{Loesch
  et~al.}{2021}]{loesch2021impermanent}
Loesch, S., N.~Hindman, M.~B. Richardson, and N.~Welch (2021).
\newblock Impermanent loss in uniswap v3.

\bibitem[\protect\citeauthoryear{Madhavan}{Madhavan}{1992}]{madhavan1992trading}
Madhavan, A. (1992).
\newblock Trading mechanisms in securities markets.
\newblock {\em the Journal of Finance\/}~{\em 47\/}(2), 607--641.

\bibitem[\protect\citeauthoryear{Malinova and Park}{Malinova and
  Park}{2023}]{malinova2023learning}
Malinova, K. and A.~Park (2023).
\newblock Learning from defi: Would automated market makers improve equity
  trading?

\bibitem[\protect\citeauthoryear{Milionis, Moallemi, and Roughgarden}{Milionis
  et~al.}{2023}]{milionis2023automated}
Milionis, J., C.~C. Moallemi, and T.~Roughgarden (2023).
\newblock Automated market making and arbitrage profits in the presence of
  fees.
\newblock {\em arXiv preprint arXiv:2305.14604\/}.

\bibitem[\protect\citeauthoryear{Milionis, Moallemi, Roughgarden, and
  Zhang}{Milionis et~al.}{2022}]{milionis2022automated}
Milionis, J., C.~C. Moallemi, T.~Roughgarden, and A.~L. Zhang (2022).
\newblock Automated market making and loss-versus-rebalancing.
\newblock {\em arXiv preprint arXiv:2208.06046\/}.

\bibitem[\protect\citeauthoryear{Nicholas and Schwartz}{Nicholas and
  Schwartz}{1995}]{nicholas1995electronic}
Nicholas, E. and R.~A. Schwartz (1995).
\newblock Electronic call market trading: Let competition increase efficiency.
\newblock {\em Journal of Portfolio Management\/}~{\em 21}, 10--18.

\bibitem[\protect\citeauthoryear{Park}{Park}{2023}]{park2022conceptual}
Park, A. (2023).
\newblock The conceptual flaws of decentralized automated market making.
\newblock {\em Management Science\/}~{\em 69\/}(11), 6731--6751.

\bibitem[\protect\citeauthoryear{Qin, Zhou, and Gervais}{Qin
  et~al.}{2022}]{qin2022quantifying}
Qin, K., L.~Zhou, and A.~Gervais (2022).
\newblock Quantifying blockchain extractable value: How dark is the forest?
\newblock In {\em 2022 IEEE Symposium on Security and Privacy (SP)}, pp.\
  198--214. IEEE.

\bibitem[\protect\citeauthoryear{Ramseyer, Goyal, Goel, and
  Mazi\`{e}res}{Ramseyer et~al.}{2023}]{goyal2022batch}
Ramseyer, G., M.~Goyal, A.~Goel, and D.~Mazi\`{e}res (2023).
\newblock Augmenting batch exchanges with constant function market makers.
\newblock {\em arXiv preprint arXiv:2210.04929\/}.

\bibitem[\protect\citeauthoryear{Roth}{Roth}{2002}]{roth2002economist}
Roth, A.~E. (2002).
\newblock The economist as engineer: Game theory, experimentation, and
  computation as tools for design economics.
\newblock {\em Econometrica\/}~{\em 70\/}(4), 1341--1378.

\bibitem[\protect\citeauthoryear{Roth and Xing}{Roth and
  Xing}{1997}]{roth1997turnaround}
Roth, A.~E. and X.~Xing (1997).
\newblock Turnaround time and bottlenecks in market clearing: Decentralized
  matching in the market for clinical psychologists.
\newblock {\em Journal of political Economy\/}~{\em 105\/}(2), 284--329.

\bibitem[\protect\citeauthoryear{Schlegel and Mamageishvili}{Schlegel and
  Mamageishvili}{2022}]{schlegel2022axioms}
Schlegel, J.~C. and A.~Mamageishvili (2022).
\newblock Axioms for constant function amms.
\newblock {\em arXiv preprint arXiv:2210.00048\/}.

\bibitem[\protect\citeauthoryear{Torres, Camino, et~al.}{Torres
  et~al.}{2021}]{torres2021frontrunner}
Torres, C.~F., R.~Camino, et~al. (2021).
\newblock Frontrunner jones and the raiders of the dark forest: An empirical
  study of frontrunning on the ethereum blockchain.
\newblock In {\em 30th USENIX Security Symposium (USENIX Security 21)}, pp.\
  1343--1359.

\end{thebibliography}
\bibliographystyle{chicago}

\end{document}